\documentclass[12pt]{article}

\usepackage{scicite}
\usepackage[utf8]{inputenc}

\usepackage{times}

\usepackage{amsmath}
\usepackage{amsfonts}
\usepackage{amssymb}
\usepackage{graphicx}

\usepackage{siunitx}
\usepackage{nameref}
\usepackage{booktabs}

\usepackage{xr}

\def\mean<#1>{\langle #1 \rangle}

\newenvironment{sciabstract}{%
\begin{quote} \bf}
{\end{quote}}

\title{Universal scaling between wave speed and size enables nanoscale high-performance reservoir computing based on propagating spin-waves
}
\author
{Satoshi Iihama,${}^{1,2}$ Yuya Koike,${}^{2,3,5}$ Shigemi Mizukami,${}^{2,4}$ Natsuhiko Yoshinaga${}^{2,5\ast}$\\
\\
\normalsize{${}^{1}$Frontier Research Institute for Interdisciplinary Sciences (FRIS), Tohoku University,}\\
\normalsize{Sendai, 980-8578, Japan}\\
\normalsize{${}^{2}$WPI Advanced Institute for Materials Research (AIMR), Tohoku University,}\\
\normalsize{Katahira 2-1-1, Sendai, 980-8577, Japan}\\
\normalsize{${}^{3}$Department of Applied Physics, Tohoku University,}\\
\normalsize{Sendai, 980-8579, Japan}\\
\normalsize{${}^{4}$Center for Science and Innovation in Spintronics (CSIS), Tohoku University,}\\
\normalsize{Sendai, 980-8577, Japan}\\
\normalsize{${}^{5}$MathAM-OIL, AIST, Sendai, 980-8577, Japan}\\
\\
\normalsize{$^\ast$To whom correspondence should be addressed; E-mail:  yoshinaga@tohoku.ac.jp.}
}

\date{}

\begin{document} 


\baselineskip24pt


\maketitle



\begin{sciabstract}
Neuromorphic computing using spin waves is promising for high-speed nanoscale devices, but the realization of high performance has not yet been achieved.
Here we show, using micromagnetic simulations and simplified theory with response functions, that spin-wave physical reservoir computing can achieve miniaturization down to nanoscales keeping high computational power comparable with other state-of-art systems.
We also show the scaling of system sizes with the propagation speed of spin waves plays a key role to achieve high performance at nanoscales.
\end{sciabstract}



\section*{Introduction}

Non-local magnetization dynamics in a nanomagnet, spin-waves, can be used for processing information in an energy-efficient manner since spin-waves carry information in a magnetic material without Ohmic losses\cite{Chumak2015}. 
The wavelength of the spin-wave can be down to the nanometer scale, and the spin-wave frequency becomes several GHz to THz frequency, which are promising properties for nanoscale and high-speed operation devices. 
Recently, neuromorphic computing using spintronics technology has attracted great attention for the development of future low-power consumption artificial intelligence\cite{grollier2020neuromorphic}.
Spin-waves can be created by various means such as magnetic field, spin-transfer torque, spin-orbit torque, voltage induced change in magnetic anisotropy and can be detected by the magnetoresistance effect\cite{barman20212021}.
Therefore, neuromorphic computing using spin waves may have a potential of realisable devices.

Reservoir computing (RC) is a promising neuromorphic computation framework.
RC is a variant of recurrent neural networks (RNNs) and  has a single layer, referred to as a reservoir, to transform an input signal into an output\cite{Jaeger:2004}.
In contrast with the conventional RNNs, RC does not update the weights in the reservoir.
Therefore, by replacing the reservoir of an artificial neural network with a physical system, for example, magnetization dynamics, we may realize a neural network device to perform various tasks, such as time-series prediction\cite{Jaeger:2004,Pathak:2018}, short-term memory\cite{Jaeger:2002,Maass:2002}, pattern recognition, and pattern generation.
Several physical RC has been proposed: spintronic oscillators\cite{Torrejon:2017,Tsunegi:2019},  optics\cite{Rafayelyan:2020}, photonics\cite{Larger:2012,Takano:2018}, fluids, soft robots, and others (see reviews \cite{Lukosevicius:2009,VanderSande:2017,Tanaka:2019}).
Among these systems, spintronic RC has the advantage in its potential realization of nanoscale devices at high speed of GHz frequency with low power consumption, which may outperform conventional electric computers in future.
So far, spintronic RC has been considered using spin-torque oscillators\cite{Torrejon:2017,Tsunegi:2019}, magnetic skyrmion\cite{Prychynenko:2018}, and spin waves in garnet thin films\cite{Nakane:2018,Ichimura:2021,Nakane:2021}. 
However, the current performance of spintronic RC still remains poor compared with the Echo State Network (ESN)\cite{Jaeger:2002,Maass:2002}, idealized RC systems.
The biggest issue is a lack of our understanding of how to achieve high performance in the RC systems.

To achieve high performance, the reservoir has to have a large degree of freedom, $N$.
However, in practice, it is difficult to increase the number of physical nodes, $N_p$,
because it requires more wiring of multiple inputs.
In this respect, wave-based computation in continuum media has attracting features.
The dynamics in the continuum media have large, possibly infinite, degrees of freedom.
In fact, several wave-based computations have been proposed\cite{Hughes:2019,Marcucci:2020}.
The challenge is to use the advantages of both wave-based computation and RC to achieve high-performance computing of time-series data.
For spin wave-based RC, so far, the large degrees of freedom are extracted only by using a large number of input and/or output nodes \cite{Nakane:2021,Dale:2021}.
Here, to propose a realisable spin wave RC, we use an alternative route; we extract the information from the continuum media using a small number of physical nodes.

Along this direction, using $N_v$ virtual nodes for the dynamics with delay was proposed to increase $N$ in \cite{Appeltant:2011}.
This idea was applied in optical fibres with a long delay line\cite{Larger:2012} and a network of oscillators with delay\cite{Rohm:2018}.
Nevertheless, the mechanism of high performance remains elusive, and no unified understanding has been made.
The increase of $N=N_p N_v$ with $N_v$ does not necessarily improve performance. 
In fact, RC based-on STO struggles with insufficient performance both in experiments\cite{Tsunegi:2019} and simulations \cite{Furuta:2018}.
The photonic RC requires a large size of devices due to the long delay line\cite{Larger:2012,Takano:2018}.

In this work, we show nanoscale and high-speed RC based on spin wave propagation with a small number of inputs can achieve performance comparable with the ESN and other state-of-art RC systems.
More importantly, by using a simple theoretical model, we clarify the mechanism of the high performance of spin wave RC.  
We show the scaling between wave speed and system size  to make virtual nodes effective.

\section*{Results}

\subsection*{Reservoir computing using wave propagation}\label{sec.concept}

\begin{figure}[htbp]%
\centering
\includegraphics[width=0.95\textwidth]{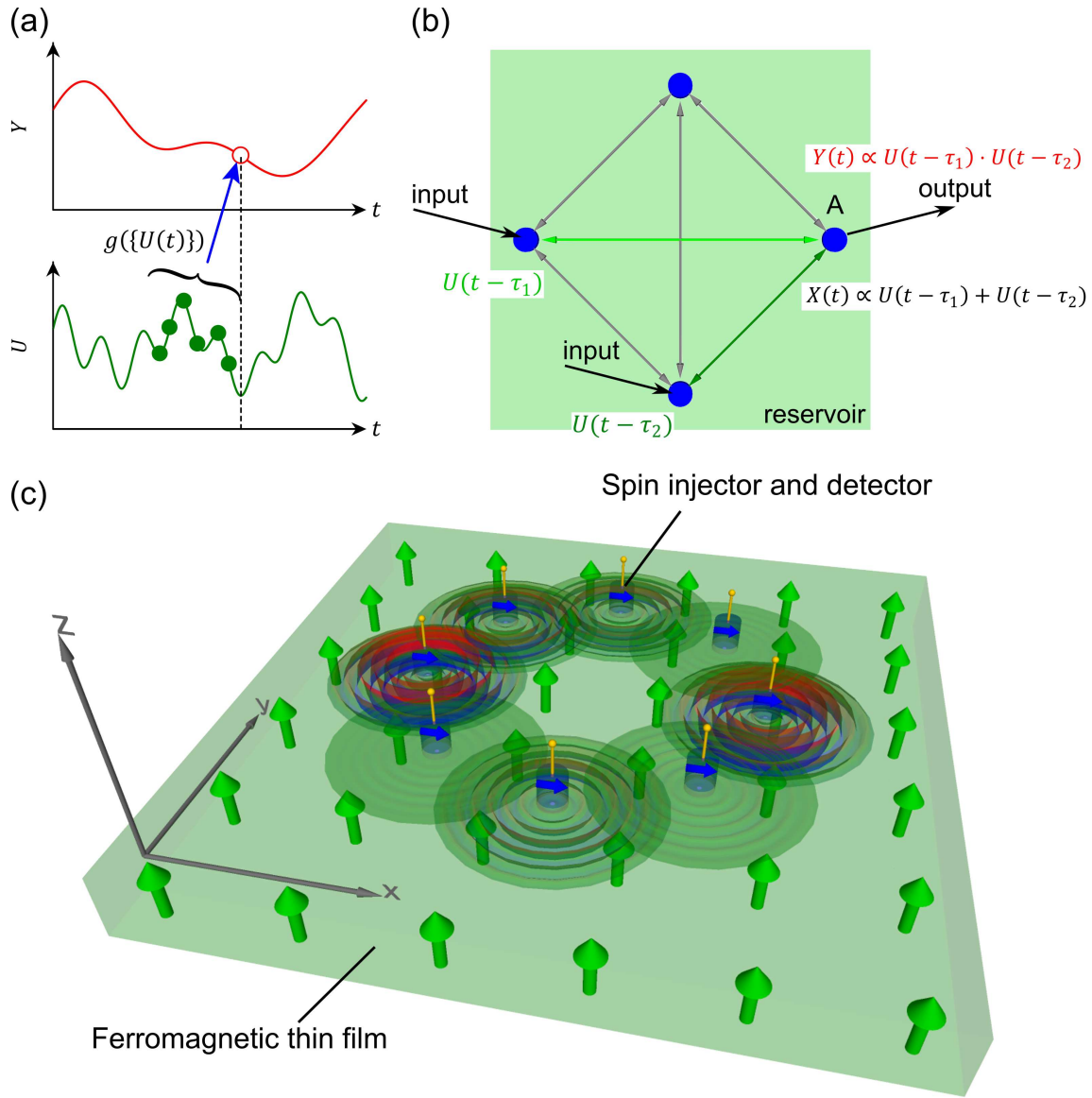}
 \caption{
 \textbf{Illustration of physical reservoir computing and reservoir based on propagating spin-wave network.} (a) Schematic illustration of output function prediction by using time-series data. Output signal $Y$ is transformed by past information of input signal $U$. (b) Schematic illustration of reservoir computing with multiple physical nodes. The output signal at physical node A contains past input signals in other physical nodes, which are memorized by the reservoir. (c) Schematic illustration of reservoir computing based on propagating spin-wave. Propagating spin-wave in ferromagnetic thin film (${\bf m} \parallel {\bf e}_{\rm z}$) is excited by spin-transfer torque at multiple physical nodes with reference magnetic layer (${\bf m} \parallel {\bf e}_{\rm x}$). x-component of magnetization is detected by the magnetoresistance effect at each physical node.
}\label{fig1}
\end{figure}

The basic task of RC is to transform an input signal $U_n$ to an output $Y_n$ for the discrete step $n=1,2,\ldots, T$ at the time $t_n$.
For example, for speech recognition, the input is an acoustic wave, and the output is a word corresponding to the sound.
Each word is determined not only by the instantaneous input but also by the past history.
Therefore, the output is, in general, a function of all the past input, $Y_n = g\left( \{ U_m \}_{m=1}^n \right)$ as in Fig.~\ref{fig1}(a).
The RC can also be used for time-series prediction by setting the output as $Y_n = U_{n+1}$\cite{Jaeger:2004}.
In this case, the state at the next time step is predicted from all the past data; namely, the effect of delay is included.
The performance of the input-output transformation $g$ can be characterized by how much past information does $g$ have, and how much nonlinear transformation does $g$ perform. 
We will discuss that the former is expressed by memory capacity (MC)\cite{Stelzer:2020}, whereas the latter is measured by information processing capacity(IPC)\cite{Dambre:2012}.

We propose physical computing based on a propagating wave (see Fig.~\ref{fig1}(b,c)).
Time series of an input signal $U_n$ can be transformed into an output signal $Y_n$ (Fig.~\ref{fig1}(a)).
As we will discuss below, this transformation requires large linear and nonlinear memories; for example, to predict $Y_n$, we need to memorize the information of $U_{n-2}$ and $U_{n-1}$.
The input signal is injected in the first input node and propagates in the device to the output node spending a time $\tau_1$ as in Fig.~\ref{fig1}(b).
Then, the output may have past information at $t_n-\tau_1$ corresponding to the step $n-m_1$.
The output may receive the information from another input at different time $t_n-\tau_2$.
The sum of the two peices of information is mixed and transformed as $U_{n-m_1} U_{n-m_2}$ either by nonlinear readout or by nonlinear dynamics of the reservoir (see also Sec.~\ref{sec.mumaxmxmz} in Supplementary Information).
We will demonstrate the wave propagation can indeed enhances memory capacity and learning performance of the input-output relationship.

Before explaining our learning strategy, we discuss how to achieve accurate learning of the input-output relationship $Y_n = g \left( \{ U_m \}_{m=1}^n \right)$ from the data.
Here, the output may be dependent on a whole sequence of the input $\{ U_m \}_{m=1}^n = \left( U_1, \ldots, U_n \right)$.
Even when both $U_n$ and $Y_n$ are one-variable time-series data, the input-output relationship $g(\cdot)$ may be $T$-variable polynomials, where $T$ is the length of the time series.
Formally, $g(\cdot)$ can be expanded in a polynomial series (Volterra series)
 such that $g \left( \{ U_m \}_{m=1}^n \right) = \sum_{k_1,k_2,\cdots,k_t} \beta_{k_1,k_2,\cdots,k_t} U^{k_1}_1 U^{k_2}_2 \cdots U^{k_n}_n$ with the coefficients $\beta_{k_1,k_2,\cdots,k_n}$.
Therefore, even for the linear input-output relationship, we need $T$ coefficients in $g(\cdot)$, and as the degree of powers in the polynomials increases, the number of the coefficients increases exponentially.
This observation implies that a large number of data is required to estimate the input-output relationship.
Nevertheless, we may expect a dimensional reduction of $g(\cdot)$ due to its possible dependence on the time close to $t$ and on the lower powers.
Still, our physical computers should have degrees of freedom $N \gg 1$, if  not exponentially large.

The reservoir computing framework is used to handle time-series data of the input $\mathbf{U}$ and the output $\mathbf{Y}$\cite{Jaeger:2002}.
In this framework, the input-output relationship is learned through the reservoir dynamics $X(t)$, which in our case, is magnetization at the detectors.
The reservoir state at a time $t_n$ is driven by the input at the $n$th step corresponding to $t_n$ as
\begin{align}
    X(t_{n+1})
    &=
    f \left(
    X(t_n),U_n
    \right)
    \label{reservoir.dynamics}
\end{align}
with nonlinear (or possibly linear) function $f(\cdot)$.
The output is approximated by the readout operator $\psi (\cdot)$ as
\begin{align}
    \hat{Y}_n
    &=
    \psi \left(
    X(t_n)
    \right)
    .
\end{align}
Our study uses the nonlinear readout $\psi \left( X(t) \right) = W_1 X(t) + W_2 X^2(t)$\cite{Pathak:2018,Bollt:2021}.
The weight matrices $W_1$ and $W_2$ are estimated from the data of the reservoir dynamics $X(t)$ and the true output $Y_n$, where $X(t)$ is obtained by \eqref{reservoir.dynamics}.
With the nonlinear readout, the RC with linear dynamics can achieve nonlinear transformation, as Fig.\ref{fig1}(b).
We stress that the system also works with linear readout when the RC has nonlinear dynamics.
We discuss this case in Sec.\ref{sec.mumaxmxmz}.

\subsection*{Spin wave reservoir computing}\label{sec.mumax}

\begin{figure}[htbp]%
\centering
\includegraphics[width=0.95\textwidth]{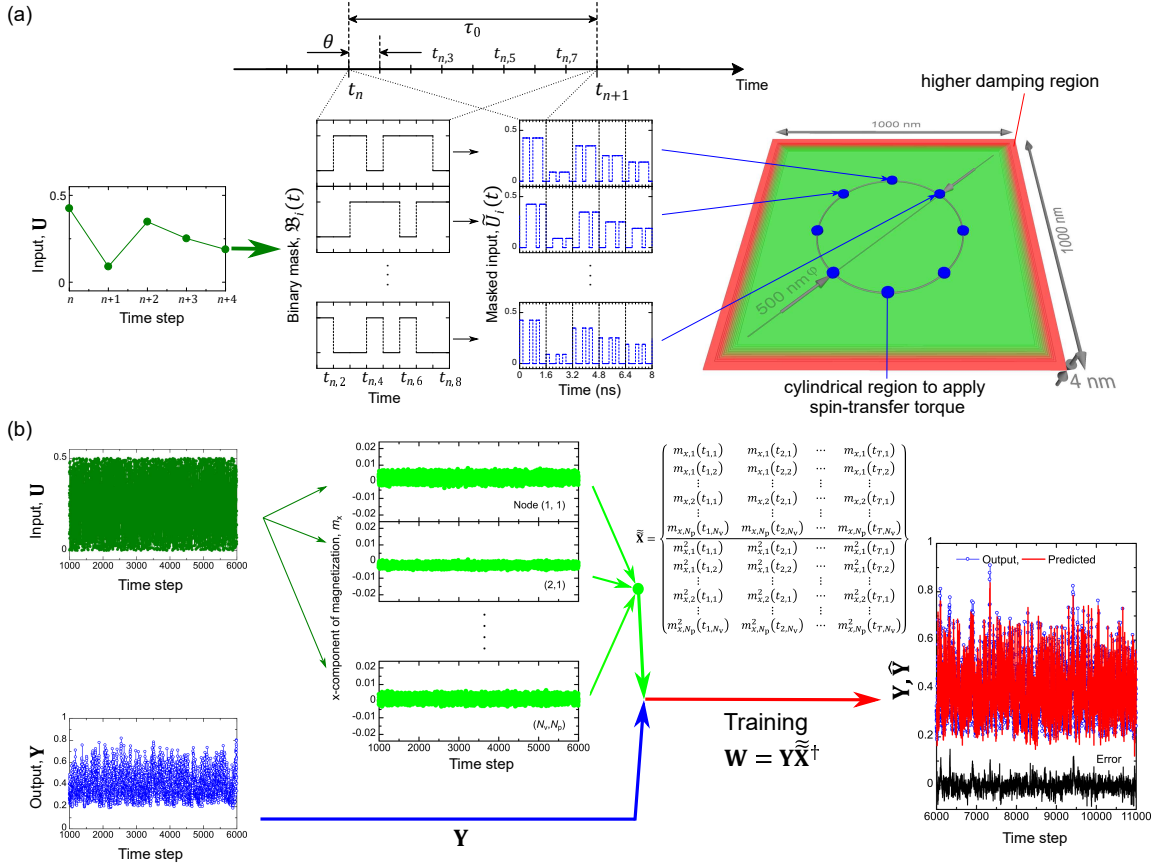}
 \caption{
 \textbf{Dimension of spin-wave reservoir and prediction of NARMA10 task.} (a) Input signals ${\bf U}$ are multiplied by binary mask $\mathcal{B}_i (t)$ and transformed into injected current $j (t) =2 j_c \tilde{U}_i (t)$ for the $i$th physical node. Current is injected into each physical node with the cylindrical region to apply spin-transfer torque and to excite spin-wave. Higher damping regions in the edges of the rectangle are set to avoid reflection of spin-waves. (b) Prediction of NARMA10 task. x-component of magnetization at each physical and virtual node are collected and output weights are trained by linear regression. 
}\label{fig2}
\end{figure}

We consider a magnetic device of a thin rectangular system with cylindrical injectors (see Fig.\ref{fig1}(c)).
The size of the device is $L \times L \times D$.
Under the uniform external magnetic field, the magnetization is along the $z$ direction. 
Electric current is injected at the $N_p$ injectors with the radius $a$ and the same height with the device.  
The spin-torque by the current drives magnetization $\mathbf{m}(\mathbf{x},t)$ and propagating spin-waves as schematically shown in Fig.\ref{fig1}(c).
The actual demonstration of the spin-wave reservoir computing is shown in Fig.~\ref{fig2}.
We demonstrate the spin-wave RC using two methods: the micromagnetic
simulations and the theoretical model using a response function.

In the micromagnetic simulations, we analyze the Landau-Lifshitz-Gilbert (LLG) equation with the effective magnetic field $\mathbf{H}_{\mathrm{eff}} = \mathbf{H}_{\mathrm{ext}} + \mathbf{H}_{\mathrm{demag}} + \mathbf{H}_{\mathrm{exch}}$ consists of the external field, demagnetization, and the exchange interaction (see \nameref{sec.method.green.function} in Methods).
The spin waves are driven by Slonczewski spin-transfer torque\cite{Slonczewski:1999}.
The driving term is proportional to the DC current $j(t)$ at the nanocontact.
We inject the DC current proportional to the input time series $\mathbf{U}$ with a pre-processing filter.
From the resulting spatially inhomogeneous magnetization $\mathbf{m}(\mathbf{x},t)$, we measure the averaged magnetization at $i$th nanocontact $\mathbf{m}_i(t)$. 
We use the method of time multiplexing with $N_v$ virtual nodes\cite{Appeltant:2011}.
We choose the $x$-component of magnetization $m_{x,i}$ as a reservoir state, namely, $X_n = \{ m_{x,i} (t_{n,k}) \}_{i\in [1,N_p],k\in [1,N_v]}$ (see \eqref{readout.mx.virtual} in Methods for its concrete form).
For the output transformation, we use $\psi(m_{i,x}) = W_{1,i} m_{i,x} + W_{2,i} m_{i,x}^2$.
Therefore, the dimension of our reservoir is $2N_p N_v$.
The nonlinear output transformation can enhance the nonlinear transformation in reservoir\cite{Pathak:2018}, and it was shown that even under the linear reservoir dynamics, RC can learn any nonlinearity\cite{Gonon:2019,Bollt:2021}.
In Sec.~\ref{sec.mumaxmxmz} in Supplementary Information, we also discuss the linear readout, but including the $z$-component of magnetization $\mathbf{X} = \left( m_{x},m_{z} \right)$.
In this case, $m_z$ plays a similar role to $m_x^2$.
The performance of the RC is measured by three tasks: MC, IPC, and NARMA10.
The weights in the readout are trained by reservoir variable $\mathbf{X}$ and the output $\mathbf{Y}$ (Fig.\ref{fig2}(b), see also Methods).

To understand the mechanism of high performance of learning by spin wave propagation, we also consider a simplified model using the response function of the spin wave dynamics.
By linearizing the magnetization around $\mathbf{m}=(0,0,1)$ without inputs, we may express the linear response of the magnetization at the $i$th readout $m_i = m_{x,i} + i m_{y,i}$ to the input as(see Methods)
\begin{align}
    m_i (t) 
    &= \sum_{j=1}^{N_p} \int dt' G_{ij}(t,t') U^{(j)}(t')
    .
    \label{m.GreenFunction}
\end{align} 
Here, $U^{(j)}(t)$ is the input time series at $j$th nanocontact.
The response function has a self part $G_{ii}$, that is, input and readout nanocontacts are the same, and the propagation part $G_{ij}$, where the distance between the input and readout nanocontacts is $\lvert \mathbf{R}_i - \mathbf{R}_j \rvert$.
We use the quadratic nonlinear readout, which has a structure 
\begin{align}
    m_i^2 (t) 
    &= \sum_{j_1=1}^{N_p} \sum_{j_2=1}^{N_p} 
    \int dt_1 \int d t_2
    G_{i j_1 j_2}^{(2)}(t,t_1,t_2) 
    U^{(j_1)}(t_1) U^{(j_2)}(t_2)
    .
    \label{m2.GreenFunction}
\end{align}
The response function of the nonlinear readout is $G_{i j_1 j_2}^{(2)}(t,t_1,t_2) \propto G_{ij_1}(t,t_1) G_{ij_2}(t,t_2)$.
The same structure as \eqref{m2.GreenFunction} appears when we use a second-order perturbation for the input (see Methods).
In general, we may include the cubic and higher-order terms of the input.
This expansion leads to the Volterra series of the output in terms of the input time series, and suggests how the spin wave RC works (see Sec.~\ref{sec.mcipc.volterra} in Supplementary Information for more details).
Once the magnetization at each nanocontact is computed, we may estimate MC and IPC.

\begin{figure}[htbp]%
\centering
\includegraphics[width=0.95\textwidth]{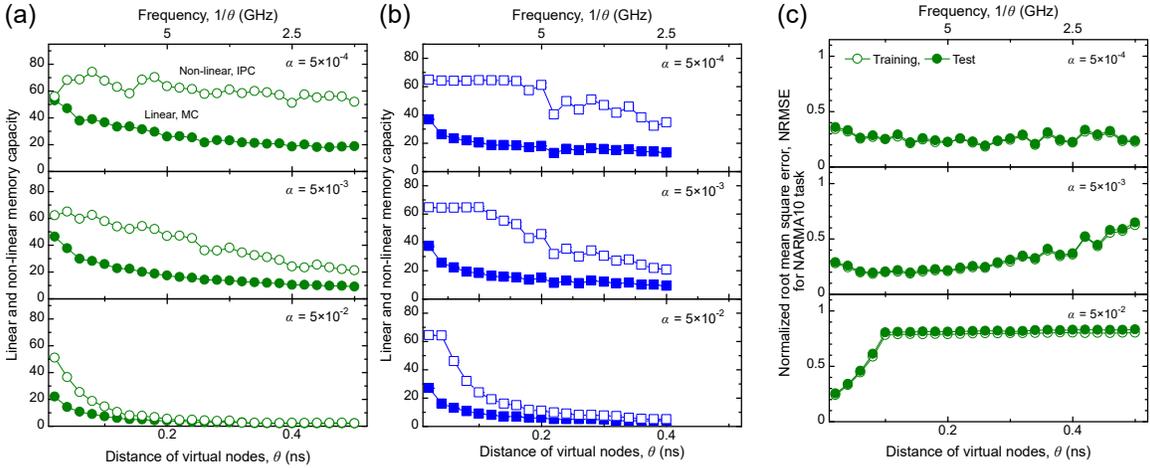}
 \caption{
 \textbf{Effect of virtual node distance on performance of spin-wave reservoir computing obtained with 8 physical nodes and 8 virtual nodes.} Memory capacity MC and information processing capacity IPC obtained by (a) micromagnetics simulation and (b) response function method plotted as a function of virtual node distance $\theta$ with different damping parameters $\alpha$.  (c) Normalized root mean square error, NRMSE for NARMA10 task is plotted as a function of $\theta$ with different $\alpha$. 
}\label{fig3}
\end{figure}

Figure~\ref{fig3} shows the results of the three tasks.
When the time scale of the virtual node $\theta$ is small and the damping is small, the performance of spin wave RC is high.
As Fig.~\ref{fig3}(a) shows, we achieve $\mathrm{MC} \approx 60$ and $\mathrm{IPC} \approx 60$.
Accordingly, we achieve a small error in the NARMA10 task, $\mathrm{NRMSE} \approx 0.2$ (Fig.~\ref{fig3}(c)).
Theses performances are comparable with state-of-the-art ESN with the number of nodes $\sim 100$.
When the damping is stronger, both MC and IPC become smaller.
Because the NARMA10 task requires the memory with the delay steps $\approx 10$ and the second order nonlinearity with the delay steps $\approx 10$ (see Sec.\ref{sec.narma10} in Supplementary Information), the NRMSE becomes larger when $\mathrm{MC} \lesssim 10$ and $\mathrm{IPC} \lesssim 10^2/2$.

The results of the micromagnetic simulations are semi-quantitatively reproduced by the theoretical model using the response function, as shown in Fig.~\ref{fig3}(b).
This result suggests that the linear response function $G(t,t')$
captures the essential feature of delay $t-t'$ due to wave propagation.

\begin{figure}[htbp]%
\centering
\includegraphics[width=0.95\textwidth]{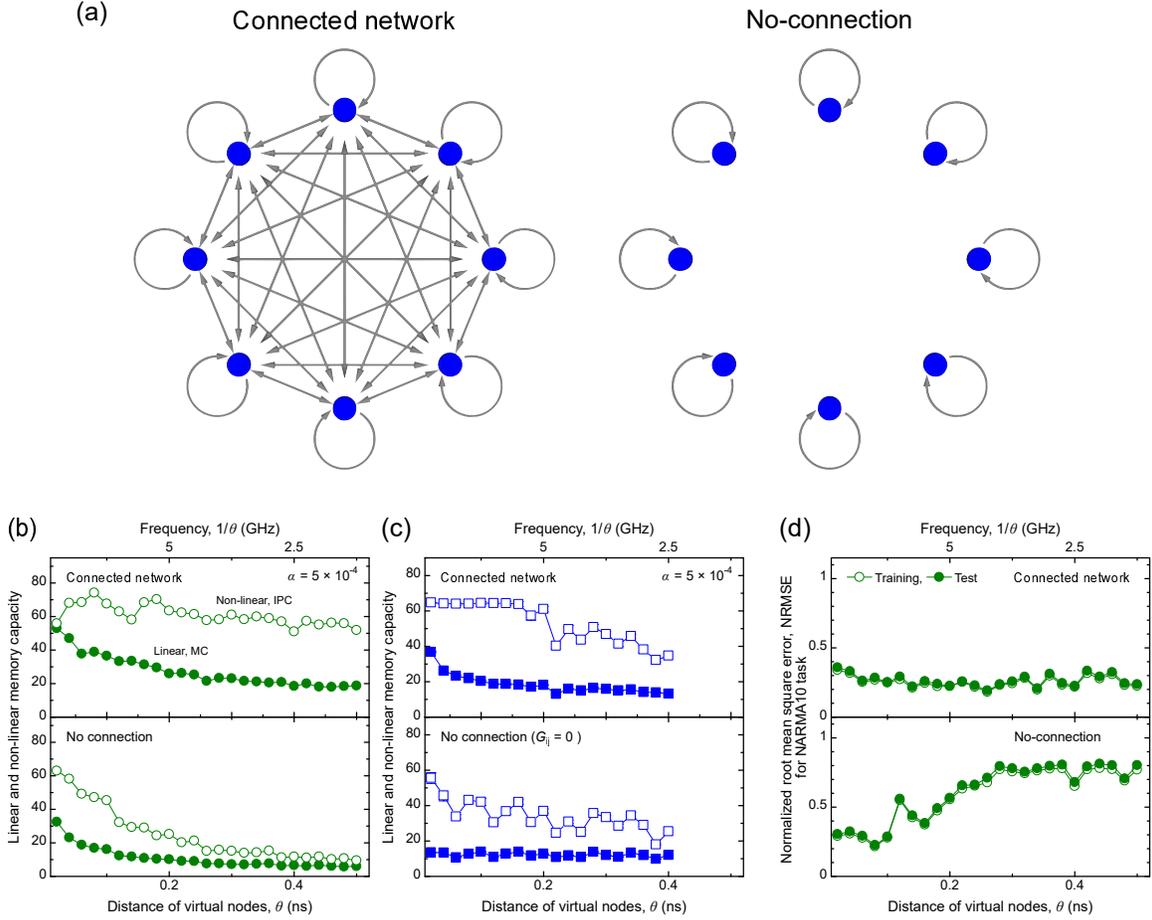}
 \caption{
 \textbf{Effect of the network connection on the performance of reservoir computing.}
(a) Schematic illustration of the network of physical nodes connected through propagating spin-wave [left] and physical nodes with no connection [right]. Memory capacity MC and information processing capacity IPC obtained using a connected network with 8 physical nodes [top] and physical nodes with no connection [bottom] calculated by (a) micromagnetics simulation and (b) response function method plotted as a function of virtual node distance $\theta $. 8 virtual nodes are used. (c) Normalized root mean square error, NRMSE for NARMA10 task obtained by micromagnetics simulation is plotted as a function of $\theta $ with a connected network [top] and physical nodes with no connection [bottom].
}\label{fig4}
\end{figure}

To confirm the high MC and IPC are due to spin-wave propagation, we perform micromagnetic simulations with damping layers between nodes (Fig.~\ref{fig4}(a)).
The damping layers inhibit spin wave propagation.
The result of Fig.~\ref{fig4}(b) shows that the memory capacity is substantially lower than that without damping, particularly when $\theta$ is small.
The NARMA10 task shows a larger error (Fig.~\ref{fig4}(d)).
When $\theta$ is small, the suppression is less effective.
This may be due to incomplete suppression of wave propagation.

We also analyze the theoretical model with the response function by neglecting the interaction between two physical nodes, namely, $G_{ij}=0$ for $i \neq j$.
In this case, information transmission between two physical nodes is not allowed.
We obtain smaller MC and IPC than the system with wave propagation,
supporting our claim (see (Fig.~\ref{fig4}(c))).

\begin{figure}[htbp]%
\centering
\includegraphics[width=0.9\textwidth]{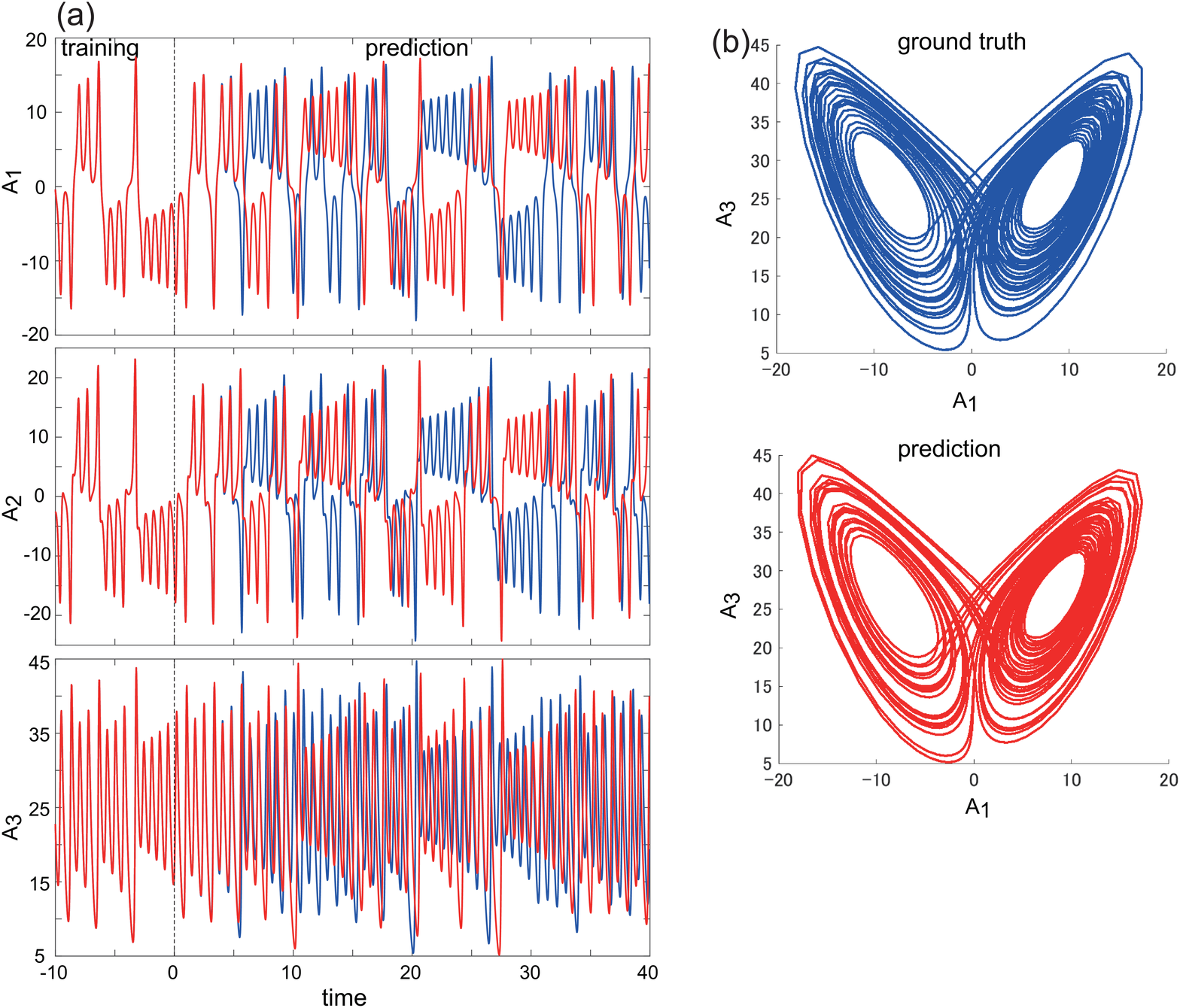}
 \caption{
 \textbf{Prediction of time-series data for the Lorenz system using the RC with micromagnetic simulations.}
The parameters are $\theta = 0.4$ns and $\alpha=5.0 \times 10^{-4}$.
(a) The ground truth $(A_1 (t),A_2 (t),A_3 (t))$ and the estimated time series $(\hat{A_1}(t),\hat{A_1}(t),\hat{A_3}(t))$ are shown in blue and red, respectively.
The training steps are during $t<0$, whereas the prediction steps are during $t>0$. 
(b) The attractor in the $A_1 A_3$ plane for the ground truth and during the prediction steps.
}\label{fig.Lorenz}
\end{figure}

Our spin wave RC also works for the prediction of time-series data.
In the study of \cite{Pathak:2018}, the functional relationship between the state at $t + \Delta t$ and the states before $t$ is learned by the ESN.
The trained ESN can estimate the state at $t + \Delta t$ from the past states, and therefore, it can predict the dynamics without the data.
In \cite{Pathak:2018}, the prediction for the chaotic time-series data was demonstrated.
Figure \ref{fig.Lorenz} shows the prediction using our spin wave RC for the Lorenz model.
We can demonstrate that the RC shows short-time prediction and, more
importantly, reconstruct the chaotic attractor.

\subsection*{Scaling of system size and wave speed}

To clarify the mechanism of the high performance of our spin wave RC, we investigate MC and IPC of the system with different characteristic length scales $L$ and different wave propagating speed $v$.
The characteristic length scale is controlled by the radius of the circle on which inputs are located (see Fig.~\ref{fig2}(a)).
We use our theoretical model with the response function to compute MC and IPC in the parameter space $(v,R)$.
This calculation can be done because the computational cost of our model is much cheaper than numerical micromagnetic simulations.

\begin{figure}[htbp]%
\centering
\includegraphics[width=0.9\textwidth]{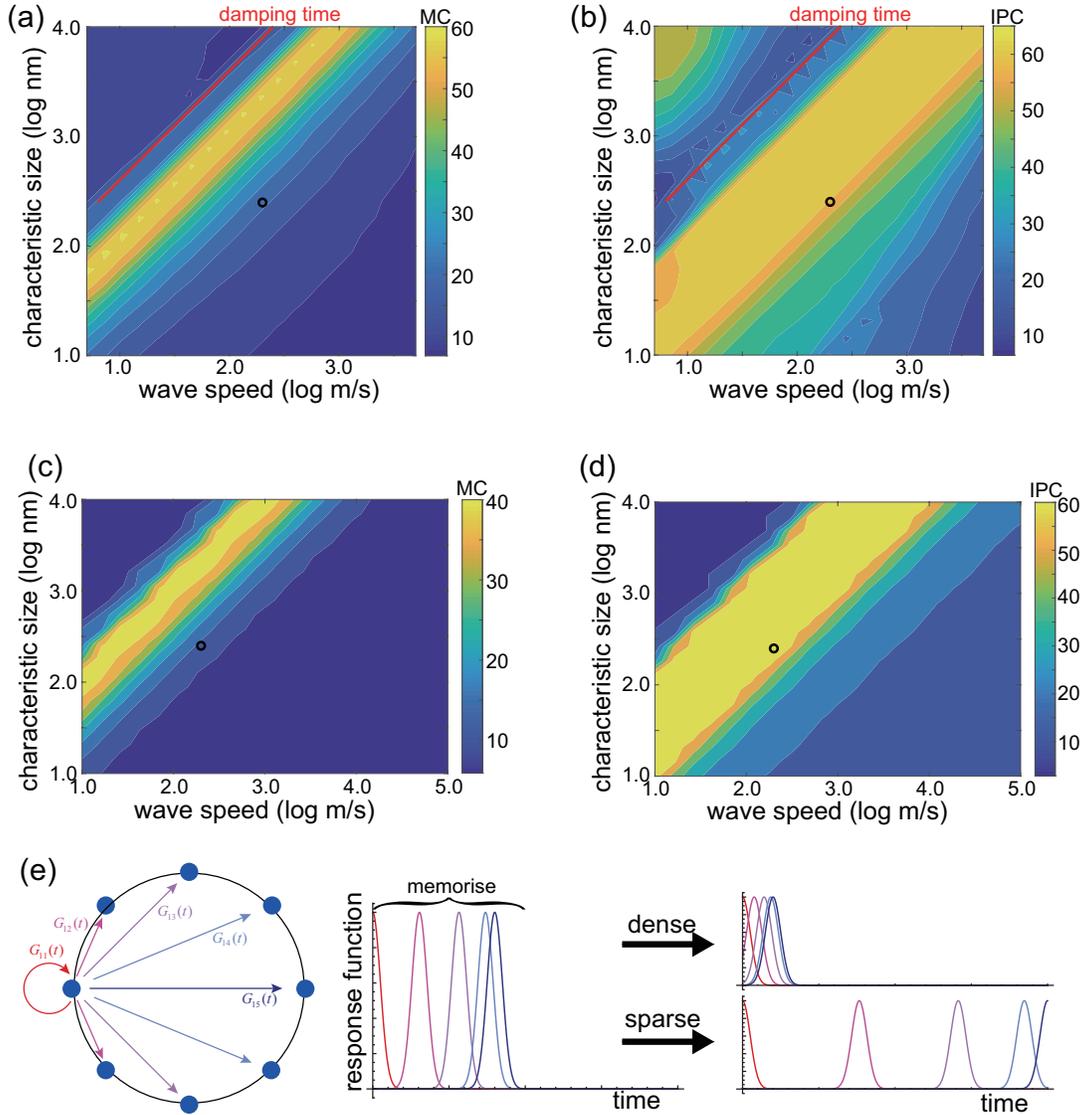}
 \caption{
 \textbf{Scaling between characteristic size and propagating wave speed obtained by response function method.}
MC (a,c) and IPC (b,d) as a function of the characteristic length scale between physical nodes $R$ and the speed of wave propagation $v$.
The results with the response function for the dipole interaction (a,b) and for the Gaussian function \eqref{Greenfunction.Gaussian} (c,d) are shown.
(e) Schematic illustration of the response function and its relation to wave propagation between physical nodes.
When the speed of the wave is too fast, all the response functions are overlapped (dense regime), while the response functions cannot cover the time windows when the speed of the wave is too slow (sparse regime). 
}\label{fig5}
\end{figure}

Figure~\ref{fig5}(a,b) shows that both MC and IPC have maximum when $L \propto v$.
To obtain a deeper understanding of the result, we perform the same analyzes for the further simplified model, in which the response function is replaced by the Gaussian function
\begin{align}
    G_{ij} (t)
    &=
    \exp \left(
    - \frac{1}{2 w^2}
    \left(
    t - \frac{R_{ij}}{v}
    \right)^2
    \right)
    \label{Greenfunction.Gaussian}
\end{align}
where $R_{ij}$ is the distance between $i$th and $j$th physical nodes, and $w$ is the width of the function. 
Even in this simplified model, we obtain MC$\approx 40$ and IPC$\approx 60$, and also the maximum when $L \propto v$ (Fig.~\ref{fig5}(c,d)).
From this result, the origin of the optimal ratio between the length and speed becomes clearer; when $L \ll v$, the response functions under different $R_{ij}$ overlap so that different physical nodes cannot carry the information of different delay times.
On the other hand, when $L \gg v$, the characteristic delay time $L/v$ exceeds the maximum delay time to compute MC and IPC, or exceeds the total length of the time series.
Note that we set the maximum delay time as 100, which is much longer
than the value necessary for the NARMA10 task.

The result suggests the universal scaling between the size of the system and the speed of the RC based on wave propagation.
Our system of the spin wave has a characteristic length $L \sim 500$ nm and a speed of $v \sim 200$ \si{\metre \per \second}.
In fact, the reported photonic RC has characteristic length scale of optical fibres close to the scaling in Fig.~\ref{fig6}.

\section*{Discussion}

\begin{figure}[htbp]%
\centering
\includegraphics[width=0.9\textwidth]{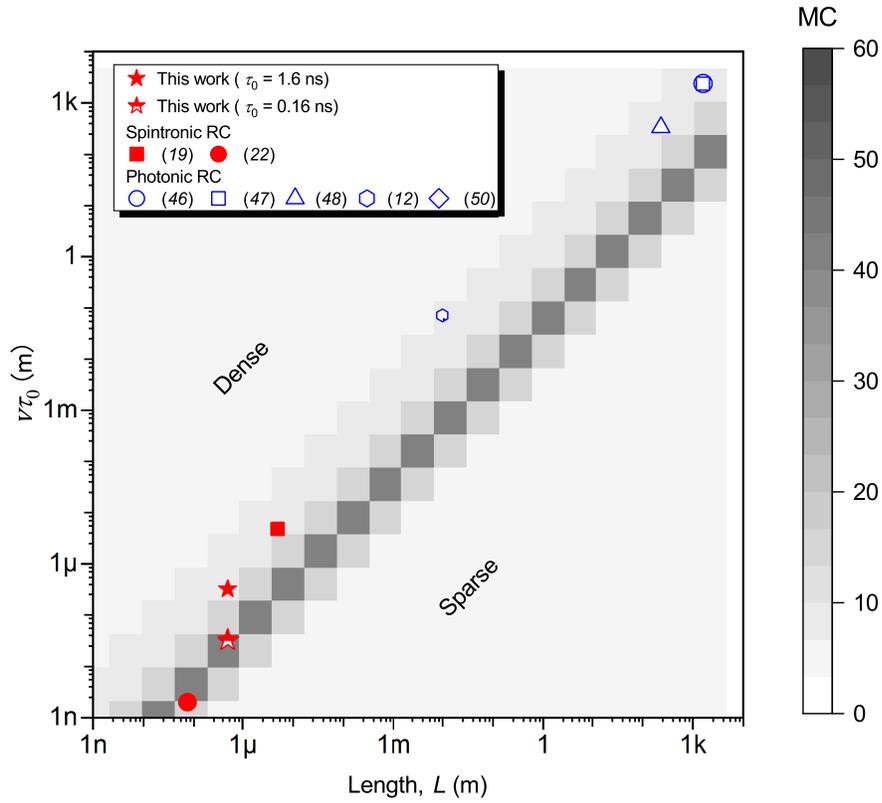}
 \caption{
 \textbf{Reports of reservoir computing using multiple nodes are plotted as a function of the length between nodes and characteristic wave speed ($v$) times delay time ($\tau _0$) for photonics system (open symbols) and spintronics system (solid symbols).}
The size of symbols corresponds to memory capacity, which is taken from literature\cite{Duport2012, Dejonckheere2014, Vinckier2015, Takano:2018, Sugano2020, Nakane:2021, Dale:2021} and this work. The gray scale represents memory capacity evaluated by using the response function method [Eq. (\ref{Greenfunction.Gaussian})].
}\label{fig6}
\end{figure}

Figure~\ref{fig6} shows reports of reservoir computing in literature with multiple nodes plotted as a function of the length of nodes $L$ and products of wave speed and delay time $v\tau _0$ for both photonic and spintronic RC. 
For the spintronic RC, the dipole interaction is considered for wave propagation in which speed is proportional to both saturation magnetization and thickness of the film \cite{Hillebrands2007}(See supplementary information sec. \ref{sec.speed}).  
For the photonic RC, the characteristic speed is the speed of light, $v \sim 10^8$ \si{\metre \per \second}.
Symbol size corresponds to MC taken from the literature [See details of plots in supplementary information sec. \ref{sec.literature}].
Plots are roughly on a broad oblique line with a ratio $L/(v\tau _0)$ $\sim $ 1.
Therefore, the photonic RC requires a larger system size, as long as the delay time of the input $\tau _0=N_v \theta $ is the same order ($\tau _0= 0.3-3$ ns in our spin wave RC). 
As can be seen in Fig.~\ref{fig5}, if one wants to reduce the length of physical nodes, one must reduce wave speed or delay time; otherwise the information is dense, and the reservoir cannot memorize many degrees of freedom (See Fig. \ref{fig5}(e)).
Reducing delay time is challenging since the experimental demonstration of the photonic reservoirs has already used the short delay close to the instrumental limit.
Also, reducing wave speed in photonics systems is challenging. 
On the other hand, the wave speed of propagating spin-wave is much lower than the speed of light and can be tuned by configuration, thickness and material parameters.
If one reduces wave speed or delay time over the broad line in Fig.~\ref{fig6}, information becomes sparse and cannot be used efficiently(See Fig. \ref{fig5}(e)).
Therefore, there is an optimal condition for high-performance RC.

\begin{figure}[htbp]%
\centering
\includegraphics[width=0.9\textwidth]{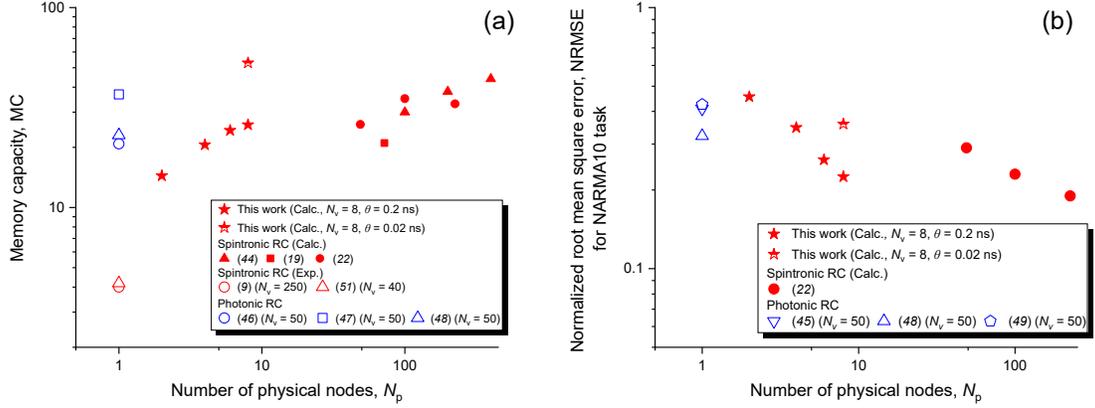}
 \caption{
 \textbf{Reservoir computing performance compared with different systems.}
(a) Memory capacity, MC reported plotted as a function of physical nodes
 $N_{\rm p}$. (b) Normalized root mean square error, NRMSE for NARMA10
 task is plotted as a function of $N_{\rm p}$. Open blue symbols are
 values reported using photonic RC while solid red symbols are values
 reported using spintronic RC. MC and NRMSE for NARMA10 task are taken from Refs. \cite{Kanao2019,Nakane:2021,Dale:2021,Tsunegi:2019, Watt2021} for spintronic RC and Refs. \cite{Duport2012, Dejonckheere2014,Vinckier2015,Duport2016, Paquot2012} for photonic RC.
}\label{fig7}
\end{figure}

The performance is comparable with other state of the art techniques, which are summarized in Fig.~\ref{fig7}.
For example, for the spintronic RC, $\mathrm{MC} \approx 30$\cite{Nakane:2021} and $\mathrm{NRMSE} \approx 0.2$\cite{Dale:2021} in the NARMA10 task are obtained using $N_p \approx 100$ physical nodes.
The spintronic RC with one physical node but with $10^1-10^2$ virtual nodes do not show high performance; MC is less than 10 (the bottom left points in Fig.~\ref{fig7}).
This fact suggests that the spintronic RC so far cannot use virtual nodes effectively.
On the other hand, for the photonic RC, comparable performances are achieved using $N_v \approx 50$ virtual nodes, but only one physical node.
As we discussed, however, the photonic RC requires mm system sizes.
Our system achieves comparable performances using $\lesssim 10$ physical nodes, and the size is down to nanoscales keeping the $2-50$ GHz computational speed.  
We also demonstrate that the spin wave RC can perform time-series prediction and reconstruction of an attractor for the chaotic data.
To our knowledge, this has not been done in nanoscale  systems.

Our results of micromagnetic simulations suggest that our system can be physically implemented.
All the parameters in this study are feasible using realistic materials\cite{Hamrle2009,kubota2009,guillemard2019,guillemard2020}.
Nanoscale propagating spin waves in a ferromagnetic thin film excited by spin-transfer torque using nanometer electrical contacts have been observed\cite{demidov2010direct,madami2011direct, sani2013mutually}. Patterning of multiple electrical nanocontacts into magnetic thin films was demonstrated in mutually synchronized spin-torque oscillators\cite{sani2013mutually}. 
In addition to the excitation of propagating spin-wave in a magnetic thin film, its non-local magnetization dynamics can be detected by tunnel magnetoresistance effect at each electrical contact, as schematically shown in Fig. \ref{fig1}(c), which are widely used for the development of spintronics memory and spin-torque oscillators. 
In addition, virtual nodes are effectively used in our system by considering the speed of propagating spin-wave and distance of physical nodes; thus, high-performance reservoir computing can be achieved with the small number of physical nodes, contrary to many physical nodes used in previous reports. 
This work provides a way to realize nanoscale high-performance reservoir
computing based on propagating spin-wave in a ferromagnetic thin film.


There is an interesting connection between our study to the recently proposed next-generation RC \cite{Gauthier:2021,Bollt:2021}, in which the linear ESN is identified with the NVAR (nonlinear vectorial autoregression) method to estimate a dynamical equation from data.
Our formula of the response function \eqref{m.GreenFunction} results in the linear input-output relationship with a delay $Y_{n+1} = a_n U_n + a_{n-1} U_{n-1} + \ldots$
(see Sec.~\ref{sec.narma10} in Supplementary Information).
More generally, with the nonlinear readout or with higher-order response functions, we have the input-output relationship with delay and nonlinearity $Y_{n+1} = a_n U_n + a_{n-1} U_{n-1} + \ldots + a_{n,n} U_n Y_n + a_{n,n-1} U_{n} U_{n-1} + \ldots$ (see Sec.~\ref{sec.mumaxmxmz} in Supplementary Information).
These input-output relations are nothing but Volterra series of the output as a function of the input with delay and nonlinearity\cite{Billings:2013}. 
The coefficients of the expansion are associated with the response function.
Therefore, the performance of RC falls into the independent components of the matrix of the response function, which can be evaluated by how much delay the response functions between two nodes cover without overlap.
The results would be helpful to a potential design of the network of the physical nodes.

We should note that the polynomial basis of the input-output relation in this study originates from spin wave excitation around the stationary state $m_z=1$.
When the input data has a hierarchical structure, another basis may be more efficient than the polynomial expansion.
Another setup of magnetic systems may lead to a different basis.
We believe that our study shows simple but clear intuition of the mechanism of high-performance RC, that can lead to the exploration of another setup for more practical application of the physical RC.


\section*{Materials and Methods}

\subsection*{Micromagnetic simulations}
We analyze the LLG equation using the micromagnetic simulator mumax$^3$\cite{Vansteenkiste:2014}.
The LLG equation for the magnetization ${\bf M} ({\bf x},t)$ yields
\begin{align}
 \partial_t {\bf M} ({\bf x},t)
 =&
 - \frac{\gamma  \mu_0}{1 + \alpha^2}
 {\bf M} \times {\bf H}_{\rm eff}
 - \frac{\alpha \gamma \mu_0}{M_s (1 + \alpha^2)}
 {\bf M} \times \left(
 {\bf M} \times {\bf H}_{\rm eff}
 \right)
 \nonumber \\
 &+ 
 \frac{\hbar P \gamma}{4 M_s^2 e  D }
 J({\bf x}, t)
 {\bf M} \times \left(
 {\bf M} \times {\bf m}_{\rm f}
 \right)
 .
 \label{LLG.mumax}
\end{align}
We consider the effective magnetic field as
\begin{align}
    \mathbf{H}_{\mathrm{eff}} 
    &= 
    \mathbf{H}_{\mathrm{ext}} + \mathbf{H}_{\mathrm{demag}} + \mathbf{H}_{\mathrm{exch}},
    \\
\mathbf{H}_{\mathrm{ext}}
&=
H_0 \mathbf{e}_z     
\\
\mathbf{H}_{\mathrm{ms}}
&=
-\frac{1}{4\pi} 
\int \nabla \nabla \frac{1}{\lvert \mathbf{r} - \mathbf{r}' \rvert} d \mathbf{r}'
\\   
\mathbf{H}_{\mathrm{exch}}
&=
\frac{2A_{\mathrm{ex}}}{\mu_0 M_s} \Delta \mathbf{M}
,
\end{align}
where $\mathbf{H}_{\mathrm{ext}}$ is the external magnetic field, $\mathbf{H}_{\mathrm{ms}}$ is the magnetostatic interaction, and $\mathbf{H}_{\mathrm{exch}}$ is the exchange interaction with the exchange parameter $A_{\mathrm{ex}}$.

The size of our system is $L=1000$ nm and $D=4$ nm.
The number of mesh points is $200$ in the $x$ and $y$ directions, and $1$ in the $z$ direction.
We consider Co$_2$MnSi Heusler alloy ferromagnet, which has a low Gilbert damping and high spin polarization with the parameter $A_{\mathrm{ex}}=23.5$ pJ/m, $M_s=1000$ kA/m, and $\alpha = 5$ $\times $ $10^{-4}$\cite{Hamrle2009, kubota2009, kubota2009, guillemard2019, guillemard2020}. Out-of-plane magnetic field $\mu _0 H_0=1.5$ T is applied so that magnetization is pointing out-of-plane.
The spin-polarized current field is included by the Slonczewski model\cite{Slonczewski:1999} with polarization parameter $P=1$ and spin torque asymmetry parameter $\lambda=1$ with the reduced Planck constant $\hbar$ and the charge of an electron $e$.
The uniform fixed layer magnetization is $\mathbf{m}_f=\mathbf{e}_x$.
We use absorbing boundary layers for spin waves to ensure the magnetization vanishes at the boundary of the system\cite{Venkat:2018}.
We set the initial magnetization as $\mathbf{m} = \mathbf{e}_z$.

The reference time scale in this system is $\tau_0 = 1/\gamma \mu_0 M_s \approx 5$ ps, where $\gamma$ is the gyromagnetic ratio, $\mu_0$ is permeability, and $M_s$ is saturation magnetization.
The reference length scale is the exchange length $l_0 \approx 5$ nm.
The relevant parameters are Gilbert damping $\alpha$, the time scale of the input time series $\theta$, and the characteristic length between the input nodes $R_0$.

The injectors and detectors of spin are placed as cylindrical nanocontacts embedded in the region with their radius $a$ and height $D$.
We set $a=20$nm unless otherwise stated.
The input time series is uniform random noise $U_n \in \mathcal{U}(0,0.5)$.
The injected density current is set as $j(t_n) = 2 j_c U_n$ with $j_c=2\times 10^{-4}/(\pi a^2)\mathrm{A/m}^2$.
Under a given input time series of the length $T$, we apply the current during the time $\theta$, and then update the current at the next step.
The same input current with different filters is injected for different virtual nodes (see \nameref{Method.learning.reservoir}).
The total simulation time is, therefore, $T \theta N_v$.

\subsection*{Learning with reservoir computing}\label{Method.learning.reservoir}

Our RC architecture consists of reservoir state variables
\begin{align}
    X(t+\Delta t)
    &=
    f \left(
    X(t), U(t)
    \right)
\end{align}
and the readout
\begin{align}
    Y_n
    &=
    \mathbf{W} \cdot \tilde{\tilde{\mathbf{X}}}(t_{n})
    .
\end{align}
In our spin wave RC, the reservoir state is chosen as $x$-component of the magnetization
\begin{align}
    \mathbf{X}
    &=
    \left(
    m_{x,1}(t_n),\ldots, m_{x,i}(t_n),\ldots, m_{x,N_p}(t_n)
    \right)^T
    ,
     \label{readout.mx}
\end{align}
for the indices for the physical nodes $i=1,2,\ldots,N_p$.
Here, $N_p$ is the number of physical nodes, and each $m_{x,i}(t_n)$ is a $T$-dimensional row vector with $n=1,2,\ldots,T$.
We use a time-multiplex network of virtual nodes in RC\cite{Appeltant:2011}, and use $N_v$ virtual nodes with time interval $\theta$.
The expanded reservoir state is expressed by $N_p N_v \times T$ matrix $\tilde{\mathbf{X}}$ as (see Fig.\ref{fig2}(b))
\begin{align}
    \tilde{\mathbf{X}}
    =&
    \left(
    m_{x,1}(t_{n,1}),
    m_{x,1}(t_{n,2}),
    \ldots,
    m_{x,1}(t_{n,k}),
    \ldots,
    m_{x,1}(t_{n,N_v}),
    \right.
    \nonumber \\
    &\left.
    \ldots,
     m_{x,i}(t_{n,1}),
    m_{x,i}(t_{n,2}),
    \ldots,
    m_{x,i}(t_{n,k}),
    \ldots,
    m_{x,i}(t_{n,N_v}),
    \ldots,
        \right.
    \nonumber \\
    &\left.
     m_{x,N_p}(t_{n,1}),
    m_{x,N_p}(t_{n,2}),
    \ldots,
    m_{x,N_p}(t_{n,k}),
    \ldots,
    m_{x,N_p}(t_{n,N_v})  
    \right)^T
    ,
    \label{readout.mx.virtual}
\end{align}
where $t_{n,k} = ( (n-1) N_v - (k-1))\theta$ for the indices of the virtual nodes $k=1,2,\ldots,N_v$.
The total number of rows is $N=N_p N_v$.
We use the nonlinear readout by augmenting the reservoir state as
\begin{align}
    \tilde{\tilde{\mathbf{X}}}
    &=
    \begin{pmatrix}
        \tilde{\mathbf{X}}\\
        \tilde{\mathbf{X}} \circ \tilde{\mathbf{X}}
    \end{pmatrix}
    ,
    \label{nonl.readout}
\end{align}
where $\tilde{\mathbf{X}}(t) \circ \tilde{\mathbf{X}}(t)$ is the Hadamard product of $\tilde{\mathbf{X}}(t)$, that is, component-wise product.
The readout weight $W$ is trained by the data of the output $Y(t)$
\begin{align}
    \mathbf{W}
    &=
     \mathbf{Y} \cdot \tilde{\tilde{\mathbf{X}}}^{\dagger}
\end{align}
where $\mathbf{X}^{\dagger}$ is pseudo-inverse of $\mathbf{X}$.

In the time-multiplexing approach, the input time-series $\mathbf{U}=(U_1,U_2, \ldots,U_T) \in \mathbb{R}^T$ is translated into piece-wise constant time-series $\tilde{U}(t) = U_n$ with $t=(n-1) N_v \theta + s$ under $k=1,\ldots,T$ and $s= [0,N_v \theta)$ (see Fig.~\ref{fig2}(a)).
This means that the same input remains during the time period $\tau_0 = N_v \theta$.
To use the advantage of physical and virtual nodes, the actual input $J_i(t)$ at the $i$th physical node is $\tilde{U}(t)$ multiplied by $\tau_0$-periodic random binary filter $\mathcal{B}_i(t)$.
Here, $\mathcal{B}_i(t) \in \{0,1\}$ is piece-wise constant during the time $\theta$.
At each physical node, we use different realizations of the binary filter as in Fig.~\ref{fig2}(a).

Unless otherwise stated, We use $1000$ steps of the input time-series as burn-in.
After these steps, we use $5000$ steps for training and $5000$ steps for
test for the MC, IPC, and NARMA10 tasks.

\subsubsection*{NARMA task}

The NARMA10 task is based on the discrete differential equation,
\begin{align}
 Y_{n+1}
 &=
 \alpha Y_n
 + \beta Y_n \sum_{p=0}^9 Y_{n-p}
 + \gamma U_n U_{n-9}
 + \delta
 .
\end{align}
Here, $U_n$ is an input taken from the uniform random distribution
$\mathcal{U}(0,0.5)$, and $y_k$ is an output.
We choose the parameter as $\alpha=0.3$, $\beta=0.05$, $\gamma=1.5$, and $\delta=0.1$.
In RC, the input is $\mathbf{U} = (U_1,U_2, \ldots, U_T)$ and the output $\mathbf{Y} = (Y_1,Y_2, \ldots, Y_T)$.
The goal of the NARMA10 task is to estimate the output time-series $\mathbf{Y}$ from the given input $\mathbf{U}$.
The training of RC is done by tuning the weights $\mathbf{W}$ so that the estimated output $\hat{Y}(t_n)$ is close to the true output $Y_n$ in terms of squared norm $\lvert \hat{Y}(t_n) - Y_n \rvert^2$.

The performance of the NARMA10 task is measured by the deviation of the estimated time series $\hat{\mathbf{Y}} = \mathbf{W} \cdot \tilde{\tilde{\mathbf{X}}}$ from the true output $\mathbf{Y}$.
The normalized root-mean-square error (NRMSE) is
\begin{align}
    {\rm NRMSE}
    & \equiv 
    \sqrt{\frac{\sum_n (\hat{Y}(t_n) - Y_n)^2}{\sum_n Y_n^2}
    }.
\end{align}
Performance of the task is high when $\mathrm{NRMSE} \approx 0$.
In the ESN, it was reported that $\mathrm{NRMSE} \approx 0.4$ for $N=50$ and $\mathrm{NRMSE} \approx 0.2$ for $N=200$\cite{Rodan:2011}.
The number of node $N=200$ was used for the speech recognition with $\approx 0.02$ word error rate\cite{Rodan:2011}, and time-series prediction of sptio-temporal chaos\cite{Pathak:2018}.
Therefore, $\mathrm{NRMSE} \approx 0.2$ is considered as reasonably high performance in practical application.
We also stress that we use the same order of nodes (virtual and physical nodes) $N=128$ to achieve $\mathrm{NRMSE} \approx 0.2$.

\subsubsection*{Memory capacity and information processing capacity}

Memory capacity (MC) is a measure of the short-term memory of RC.
This was introduced in \cite{Jaeger:2002}.
For the input $U_n$ of random time series taken from the uniform distribution, the network is trained for the output $Y_n = U_{n-k}$.
 The MC is computed from
 \begin{align}
  {\rm MC}_k
  &=
  \frac{\langle U_{n-k},\mathbf{W} \cdot \mathbf{X}(t_n) \rangle^2}{\langle U_n^2\rangle
  \langle (\mathbf{W} \cdot \mathbf{X}(t_n) )^2 \rangle}
 .
 \end{align}
 This quantity is decaying as the delay $k$ increases, and MC is defined as
 \begin{align}
  {\rm MC}
  &=
  \sum_{k=1}^{k_{\rm max}} {\rm MC}_k
  .
 \end{align}
 Here, $k_{\rm max}$ is a maximum delay, and in this study we set it as $k_{\rm max}=100$.
 The advantage of MC is that when the input is independent and
 identically distributed (i.i.d.), and the output function is linear,
 then MC is bounded by $N$, the number of internal nodes.
 
 Information processing capacity (IPC) is a nonlinear version of MC \cite{Dambre:2012}.
 In this task, the output is set as
 \begin{align}
     Y_n
     &=
     \prod_{k} \mathcal{P}_{d_k}(U_{n-k})
 \end{align}
 where $d_k$ is non-negative integer, and $\mathcal{P}_{d_k}(x)$ is the Legendre polynomials of $x$ order $d_k$.
 We may define 
  \begin{align}
  {\rm IPC}_{d_0,d_1, \ldots, d_{T-1}}
  &=
  \frac{\langle Y_n, \mathbf{W} \cdot \mathbf{X}(t_n) \rangle^2}{\langle Y_n^2\rangle
  \langle (\mathbf{W} \cdot \mathbf{X}(t_n) )^2 \rangle}
 .
 \end{align}
 and then compute $j$th order IPC as
  We may define 
  \begin{align}
  {\rm IPC}_{j}
  &=
\sum_{d_k {\rm s.t.} j = \sum_k d_k}
  {\rm IPC}_{d_1,d_2, \ldots, d_{T}}
 \label{IPC}
 .
 \end{align}
 When $j=1$, the IPC is, in fact, equivalent to MC, because $\mathcal{P}_{0}(x)=1$ and $\mathcal{P}_{1}(x)=x$.
 In this case, $Y_n=U_{n-k}$ for $d_i=1$ when $i=k$ and $d_i=0$ otherwise.
 \eqref{IPC} takes the sum over all possible delay $k$, which is nothing but MC.
 When $j>1$, IPC captures all the nonlinear transformation and delays up to the $j$th polynomial order.
 For example, when $j=2$, the output can be $Y_n = U_{n-k_1} U_{n-k_2}$ or $Y_n = U_{n-k}^2+ {\rm const.}$
 In this study, we focus on $j=2$ because the second-order nonlinearity is essential for the NARMA10 task (see Sec.~\ref{sec.narma10} in Supplementary Information).
 
 The relevance of MC and IPC is clear by considering the Volterra series of the input-output relation,
 \begin{align}
     Y_n
     &= \sum_{k_1,k_2,\cdots,k_t} \beta_{k_1,k_2,\cdots,k_n} U^{k_1}_1 U^{k_2}_2 \cdots U^{k_n}_n
     .
 \end{align}
 Instead of polynomial basis, we may use orthonormal basis such as the Legendre polynomials
  \begin{align}
     Y_n
     &= \sum_{k_1,k_2,\cdots,k_n} \beta_{k_1,k_2,\cdots,k_n} 
     \mathcal{P}_{k_1}(U_1) 
     \mathcal{P}_{k_2}(U_2)  \cdots 
     \mathcal{P}_{k_n}(U_n) 
     .
     \label{volterra.legendre}
 \end{align}
Each term in \eqref{volterra.legendre} is characterized by the non-negative indices $(k_1, k_2, \ldots, k_n)$.
Therefore, the terms corresponding to $j=\sum_i k_i=1$ in $Y_n$ have information on linear terms with time delay.
Similarly, the terms corresponding to $j=\sum_i k_i=2$ have information of second-order nonlinearity with time delay.
In this view, the estimation of the output $Y(t)$ is nothing but the estimation of the coefficients $\beta_{k_1,k_2,\ldots,k_n}$.
In RC, the readout of the reservoir state at $i$th node (either physical or virtual node) can also be expanded as the Volterra series
 \begin{align}
     \tilde{\tilde{X}}^{(i)}(t_n)
     &= \sum_{k_1,k_2,\cdots,k_n} \tilde{\tilde{\beta}}^{(i)}_{k_1,k_2,\cdots,k_n} U^{k_1}_1 U^{k_2}_2 \cdots U^{k_n}_n
     .
 \end{align}
 Therefore, MC and IPC are essentially a reconstruction of $\beta_{k_1,k_2,\cdots,k_n}$ from $\tilde{\tilde{\beta}}^{(i)}_{k_1,k_2,\cdots,k_n}$ with $i \in [1,N]$.
 This can be done by regarding $\beta_{k_1,k_2,\cdots,k_n}$ as a $T+T (T-1)/2+\cdots$-dimensional vector, and using the matrix $M$ associated with the readout weights as
 \begin{align}
     \beta_{k_1,k_2,\cdots,k_n}
     &=
     M \cdot 
     \begin{pmatrix}
     \tilde{\tilde{\beta}}^{(1)}_{k_1,k_2,\cdots,k_n}\\
     \tilde{\tilde{\beta}}^{(2)}_{k_1,k_2,\cdots,k_n}\\
     \vdots\\
     \tilde{\tilde{\beta}}^{(N)}_{k_1,k_2,\cdots,k_n}
     \end{pmatrix}
     .
 \end{align}
 MC corresponds to the reconstruction of $\beta_{k_1,k_2,\cdots,k_n}$ for $\sum_i k_i=1$, whereas the second-order IPC is the reconstruction of $\beta_{k_1,k_2,\cdots,k_n}$ for $\sum_i k_i=2$.
 If all of the reservoir states are independent, we may reconstruct $N$ components in $\beta_{k_1,k_2,\cdots,k_n}$.
 In realistic cases, the reservoir states are not independent, and therefore, we can estimate only $<N$ components in $\beta_{k_1,k_2,\cdots,k_n}$.

\subsubsection*{Prediction of chaotic time-series data}
Following \cite{Pathak:2018}, we perform the prediction of time-series data from the Lorenz model.
The model is a three-variable system of $(A_1(t),A_2(t),A_3(t))$ yielding the following equation
\begin{align}
\frac{d A_1}{dt}
&=
10 (A_2-A_1)
\\
\frac{d A_2}{dt}
&=
A_1 (28-A_3) - A_2
\\
\frac{d A_3}{dt}
&=
A_1 A_2 - \frac{8}{3}A_3
.
\end{align}
The parameters are chosen such that the model exhibits chaotic dynamics. 
Similar to the other tasks, we apply the different masks of binary noise for different physical nodes, 
$\mathcal{B}^{(l)}_i(t)$ $\in$ $\{-1,1\}$.
Because the input time series is three-dimensional, we use three independent masks for $A_1$, $A_2$, and $A_3$, therefore, $l \in \{ 1,2,3 \}$. 
The input for the $i$th physical node after the mask is given as $\mathcal{B}_i(t) \tilde{U}_i(t) = \mathcal{B}^{(1)}_i(t) A_1 (t) + \mathcal{B}^{(2)}_i(t) A_2 (t) + \mathcal{B}^{(3)}_i(t)  A_3(t)$.
Then, the input is normalized so that its range becomes $[0,0.5]$, and applied as an input current.
Once the input is prepared, we may compute magnetization dynamics for each physical and virtual node, as in the case of the NARMA10 task.
We note that here we use the binary mask of $\{-1,1 \}$ instead of $\{ 0,1 \}$ used for other tasks.
We found that the $\{0,1 \}$ does not work for the prediction of the Lorenz model, possibly because of the symmetry of the model.

The ground-truth data of the Lorenz time-series is prepared using the Runge-Kutta method with the time step $\Delta t=0.025$.
The time series is $t \in [-60,75]$, and $t \in [-60,-50]$ is used for relaxation, $t \in (-50,0]$ for training, and $t \in (0,75]$ for prediction.
During the training steps, we compute the output weight by taking the output as $\mathbf{Y} = (A_1(t+\Delta t),A_2(t+\Delta t),A_3(t+\Delta t))$.
After training, the RC learns the mapping $(A_1(t),A_2(t),A_3(t)) \rightarrow (A_1(t+\Delta t),A_2(t+\Delta t),A_3(t+\Delta t))$.
For the prediction steps, we no longer use the ground-truth input but the estimated data $(\hat{A_1}(t),\hat{A_2}(t),\hat{A_3}(t))$.
Using the fixed output weights computed in the training steps, the time evolution of the estimated time-series $(\hat{A_1}(t),\hat{A_2}(t),\hat{A_3}(t))$ is computed by the RC.

\subsection*{Theoretical analysis using response function}\label{sec.method.green.function}

We consider the Landau-Lifshitz-Gilbert equation for the magnetization field $\mathbf{m}(\mathbf{x},t)$,
\begin{align}
 \partial_t {\bf m} ({\bf x},t)
 &=
 - {\bf m} \times {\bf h}_{\rm eff}
 - {\bf m} \times \left(
 {\bf m} \times {\bf h}_{\rm eff}
 \right)
 + \sigma ({\bf x}, t)
 {\bf m} \times \left(
 {\bf m} \times {\bf m}_{\rm f}
 \right)
 \label{LLG.th}
\end{align}
We normalize both the magnetic and effective fields  by saturation
magnetization as ${\bf m} = {\bf M}/M_s$ and ${\bf h}_{\rm eff} = {\bf
H}_{\rm eff}/M_s$.
This normalization applies to all the fields including external and
anisotropic fields.
We also normalize the current density as $\sigma(\mathbf{x},t) = J(\mathbf{x},t)/j_0$ for the current density $J (\mathbf{x})$ and the unit of current density $j_0 = \frac{4 M_s^2 e \pi a^2 D \mu_0}{\hbar P}$.
We apply the current density at the nanocontact as
\begin{align}
    J (\mathbf{x},t)
    &=
    2 j_c 
    \tilde{U}(t) 
    \sum_{i=1}^{N_p}
    \chi_a (\lvert \mathbf{x} - \mathbf{R}_i \rvert)
\end{align}
Here $\chi_a (x)$ is a characteristic function $\chi_a(x)=1$ when $x \leq a$ and  $\chi_a(x)=0$ otherwise. 

We expand the solution of \eqref{LLG.th} around the uniform magnetization $\mathbf{m}(\mathbf{x},t)=(0,0,1)$ without current injection as
\begin{align}
    \mathbf{m}(\mathbf{x},t)
    &=
    \mathbf{m}^{0}(\mathbf{x},t)
    + \epsilon \mathbf{m}^{(1)}(\mathbf{x},t)
    + \mathcal{O}(\epsilon^2)
    .
    \label{m.expansion}
\end{align}
Here, $\mathbf{m}^{0}(\mathbf{x},t)= (0,0,1)$ and $\epsilon \ll 1$ is a small parameter corresponding to the magnitude of the input $\sigma (\mathbf{x},t)$.
The first-order term corresponds to a linear response of the magnetization to the input $\sigma$, whereas the higher-order terms describe nonlinear responses, for example, $\mathbf{m}^{(2)}(\mathbf{x},t) \sim \sigma(\mathbf{x}_1,t_1) \sigma(\mathbf{x}_2,t_2)$.
Because our input is driven by the spin torque with fixed layer magnetization in the $x$-direction, $\mathbf{m}_f = \mathbf{e}_x$, only $m_x$ and $m_y$ appear in the first-order term $\mathcal{O}(\epsilon)$.
Deviation of $m_z$ from $m_z=1$ appears in $\mathcal{O}(\epsilon^2)$.
Therefore, for the first-order term $\mathbf{m}^{(1)}$, we may define the complex magnetization
\begin{align}
    m
    &=
    m_x + i m_y
    .
\end{align}

Here, we will show the magnetization is expressed by the response function $G_{ij}(t)$.
The input at the $j$th physical node affects the magnetization at the $i$th physical node as
\begin{eqnarray}
 m_i(t)
  &=
  \int d\tau
  G_{ii}(t-\tau) \sigma_i (\tau)
  +
  \sum_{i \neq j}
  \int d \tau
  G_{ij}(t-\tau) \sigma_j (\tau)
  .
  \label{green.function}
\end{eqnarray}
The input for the $j$th physical node is expressed by $\sigma_j(t) = 2 j_c \mathcal{B}_j(t) \tilde{U}_j(t)$.
Because different physical nodes have different masks discussed in \nameref{Method.learning.reservoir} in Methods. 
When the wave propagation is dominated by the exchange interaction, the response function for the same node is
\begin{align}
 G_{ii} (t-\tau)
 &=
 \frac{1}{2\pi}
 e^{-\tilde{h} (\alpha + i) (t - \tau)}
 \left(
 1 -
 e^{-\frac{a^2}{4 (\alpha + i)(t - \tau)}}
 \right)
  \label{Giiexchange}
\end{align}
and for different nodes, it becomes
\begin{align}
 G_{ij} (t-\tau)
 &=
 \frac{a^2}{2\pi}
 e^{-\tilde{h} (\alpha + i) (t - \tau)}
 e^{-\frac{\lvert {\bf R}_i - {\bf R}_j \rvert^2}{4 (\alpha + i)(t - \tau)}}
 \frac{1}{2(\alpha + i)(t - \tau)}
 .
 \label{Gijexchange}
\end{align}
When the wave propagation is dominated by the dipole interaction, the response function for the same node is
\begin{align}
 G_{ii} (t-\tau)
 &=
 \frac{1}{2\pi}
 e^{-\tilde{h} (\alpha + i) (t - \tau)}
 \frac{-1 + \sqrt{1+\frac{a^2}{(d/4)^2(\alpha + i)^2(t-\tau)^2}}}{\sqrt{1+\frac{a^2}{(d/4)^2(\alpha + i)^2(t-\tau)^2}}}
 \label{Giidipole}
\end{align}
and for different nodes it becomes
\begin{align}
 G_{ij} (t-\tau)
 =&
 \frac{a^2}{2\pi}
 e^{-\tilde{h} (\alpha + i) (t - \tau)}
 \nonumber \\
 & \times
 \frac{1}{(d/4)^2 (\alpha+i)^2 (t-\tau)^2
 \left(1+\frac{\lvert {\bf R}_i - {\bf R}_j \rvert^2}{(d/4)^2 (\alpha+i)^2 (t-\tau)^2}\right)^{3/2}}
 .
  \label{Gijdipole}
\end{align}
Clearly, $G_{ii}(0) \rightarrow 1$ and $G_{ij}(0) \rightarrow 0$, while $G_{ii}(\infty) \rightarrow 0$ and $G_{ij}(\infty) \rightarrow 0$.

Once the magnetization is expressed in the form of \eqref{green.function}, we may compute the reservoir state $\mathbf{X}$ under the input $\mathbf{U}$.
Then, we may use the same method as in \nameref{Method.learning.reservoir}, and estimate the output $\hat{\mathbf{Y}}$.
Similar to the micromagnetic simulations, we evaluate the performance by MC, IPC, and NARMA10 tasks. 

We may extend the analyzes for the higher-order terms in the expansion of \eqref{m.expansion}.
In Sec.\ref{sec.mumaxmxmz} in Supplementary Materials, we show the second-order term $\mathbf{m}^{(2)} (\mathbf{x},t)$ has only the $z$-component, and moreover, it is dependent only on the first-order terms. 
As a result, the second-order term is expressed as
\begin{align}
    m_z^{(2)} (\mathbf{x},t) 
    &= 
    - \frac{1}{2} \left( 
    (m_x^{(1)})^2 + (m_y^{(1)})^2
    \right)
    .
\end{align}

To compute the response functions, we linearize \eqref{LLG.th} for the complex magnetization $m(\mathbf{x},t)$ as
\begin{align}
 \partial_t m ({\bf x},t)
 &=
 \mathcal{L} m
 + \sigma ({\bf x}, t),
\end{align}
where the linear operator is expressed as
\begin{align}
 \mathcal{L}
 &=
 \left(
-\tilde{h} + \Delta
 \right)
 \left(
\alpha + i
 \right)
 .
\end{align}
In the Fourier space, the linearized equation becomes
\begin{align}
 \partial_t m_{\bf k} (t)
 &=
 \mathcal{L}_{k} m_{\bf k}
 + \sigma_{\bf k}(t),
 \label{LLG.linear.k}
\end{align}
with
\begin{align}
 \mathcal{L}_k
 &=
- \left(
\tilde{h} + k^2
 \right)
 \left(
\alpha + i
 \right)
 .
\end{align}
The solution of (\eqref{LLG.linear.k}) is obtained as
\begin{align}
 m_{\bf k} (t)
 &=
 \int d \tau
 e^{\mathcal{L}_k (t-\tau)}
 \sigma_k (\tau)
 .
\end{align}
We have $N_p$ cylindrical shape inputs with radius $a$ and the $i$th input is located at $\mathbf{R}_i$.
The input function is expressed as
\begin{align}
    \sigma (\mathbf{x})
    &=
    \sum_{i=1}^{N_p}
    \chi_a \left(
    \lvert \mathbf{x} - \mathbf{R}_i \rvert 
    \right)
    .
\end{align}
We are interested in the magnetization at the input $m_i(t) = m(\mathbf{x}=\mathbf{R}_i,t)$, which is
\begin{align}
 {\bf m}_i
 =&
 \frac{1}{(2\pi)^2}
 \sum_{j}
 \int d \tau
 e^{-\tilde{h}(\alpha + i)(t-\tau)}
 \int dk
 e^{-k^2(\alpha + i)(t-\tau)}
 e^{i {\bf k}\cdot ({\bf R}_i - {\bf R}_j)}
 2 \pi a
 J_1 (ka)
 \sigma_j (t)
 \nonumber \\
 =&
 \frac{a}{2\pi}
 \sum_{j}
 \int d \tau
 e^{-\tilde{h}(\alpha + i)(t-\tau)}
 \int dk
 e^{-k^2(\alpha + i)(t-\tau)}
 J_0 \left(
k \lvert 
{\bf R}_i - {\bf R}_j 
\rvert
 \right)
 J_1 (ka)
 \sigma_j(t)
\end{align}
For the same node, $\lvert \mathbf{R}_i - \mathbf{R}_j \rvert=0$, and we may compute the
integral explicitly as \eqref{Giiexchange}.
When $ka \ll 1$, we may assume $J_1 (ka) \approx ka/2$, and finally, come up with \eqref{Gijexchange}.

When the thickness $d$ of the material is thin, the dispersion relation becomes
\begin{align}
 \mathcal{L}_k
 &=
 - \tilde{h} (\alpha + i)
 \sqrt{
\left(
 1 + \frac{k^2}{\tilde{h}}
 \right)
\left(
 1 + \frac{k^2}{\tilde{h}}
 + \frac{\beta k}{\tilde{h}}
 \right)
 }
\end{align}
where
\begin{align}
 \beta
 &=
 \frac{d}{2}
 .
\end{align}
We assume for $k \ll \beta \sqrt{\tilde{h}}$, then the linearized operator becomes
\begin{align}
 \mathcal{L}_k
 &=
 - (\alpha + i)
 \left(
\tilde{h} + \frac{k d}{4}
 \right)
\end{align}
leading to \eqref{Giidipole} and \eqref{Gijdipole}.


\noindent \textbf{Acknowledgements:} 
%
\\
 S. M. thanks to CSRN at Tohoku University.
  Numerical simulations in this work were carried out in part by AI
 Bridging Cloud Infrastructure (ABCI) at National Institute of Advanced
  Industrial Science and Technology (AIST), and by the supercomputer
  system at the information initiative center, Hokkaido University, Sapporo, Japan.\\
  \vskip\baselineskip
\noindent \textbf{Funding:} \\
This work is support by JSPS KAKENHI grant numbers 21H04648, 21H05000 to S.M., by JST, PRESTO Grant Number JPMJPR22B2 to S.I., X-NICS, MEXT Grant Number JPJ011438 to S.M., and by JST FOREST Program Grant Number JPMJFR2140 to
 N.Y.
 \\
 \vskip\baselineskip
\noindent \textbf{Author Contributions} \\
S.M., N.Y., S.I. conceived the research. S.I., Y.K., N.Y. carried out simulations.
N.Y., S.I. analyzed the results.
N.Y., S.I., S.M. wrote the manuscript.
All the authors discussed the results and analysis.
\\
\vskip\baselineskip
\noindent \textbf{Competing Interests}\\
The authors declare that they have no competing financial interests.\\
\vskip\baselineskip
\noindent \textbf{Data and materials availability:} \\
All data are available in the main text or the supplementary materials.



\appendix

\section{Connection between the NARMA10 task and MC/IPC}\label{sec.narma10}

In this section, we discuss the necessary properties of reservoir
computing to achieve high performance of the NARMA10 task.
In short, the NARMA10 task is dominated by the memory of nine step previous
data and second-order nonlinearity.
We discuss these properties in two methods.
The first method is based on the extended Dynamic Mode Decomposition
(DMD)\cite{Li2017} and the higher-order DMD\cite{LeClainche:2017}.
The second method is a regression of the input-output relationship.
 We will discuss the details of the two methods.
 Our results are consistent with previous studies; the requirement of memory was discussed in \cite{Carroll:2022}, and the
second-order nonlinear terms with a time delay in \cite{Kubota:2019}.

The NARMA10 task is based on the discrete differential equation,
\begin{align}
 Y_{n+1}
 &=
 \alpha Y_n
 + \beta Y_n \sum_{i=0}^9 Y_{n-i}
 + \gamma U_n U_{n-9}
 + \delta
 .
\end{align}
Here, $U_n$ is an input at the time step $n$ taken from the uniform random distribution
$\mathcal{U}(0,0.5)$, and $Y_n$ is an output.
We choose the parameter as $\alpha=0.3$, $\beta=0.05$, $\gamma=1.5$, and $\delta=0.1$.

In the first method, we estimate the transition matrix ${\bf A}$ from the
state variable ${\bf Y}_n = (Y_1, Y_2, \ldots,Y_n)$ to ${\bf Y}_{n+1} = (Y_2, Y_3, \ldots,Y_{n+1})$ yielding 
\begin{align}
 {\bf Y}_{n+1}
 &=
 {\bf A} \cdot  {\bf Y}_{n}
 .
\end{align}
We may extend the notion of the state variable to contain delayed data and polynomials of the output
with time delay as
\begin{align}
 {\bf Y}_n
 &=
 \left(
 Y_{n}, Y_{n-1}, \ldots, Y_{1},
 Y_{n} Y_{n}, Y_{n} Y_{n-1}, \ldots, Y_1 Y_1
 \right)
 .
 \label{output.extended.dmd}
\end{align}
Including the delay terms following from the higher-order DMD\cite{LeClainche:2017}, while the polynomial nonlinear terms are used as a polynomial dictionary in the extended DMD\cite{Li2017}. 
Here, \eqref{output.extended.dmd} contains all the combination of the second-order terms with time delay, $Y_{n-i_1} Y_{n-i_2}$ with the integers $i_1$ and $i_2$ in $0 \leq i_1 \leq l_2 \leq n-1$.
We may straightforwardly include higher-order terms in powers in \eqref{output.extended.dmd}.
In the NARMA10 task, the output $\mathbf{Y}_{n+1}$ is also affected by the input $\mathbf{U}_n$.
Therefore, the extended DMD is generalized to include the control as \cite{Brunton:2019}
\begin{align}
 {\bf Y}_{n+1}
 &=
\left(
 {\bf A} \;
 {\bf B}
\right)
 \cdot
\begin{pmatrix}
 {\bf Y}_{n} \\
 {\bf U}_{n}
\end{pmatrix}
 ,
 \label{narma10.exDMD}
\end{align}
where the state variable corresponding to the input includes time delay and
nonlinearity, and is described as
\begin{align}
 {\bf U}_n
 &=
 \left(
 U_{n}, U_{n-1}, \ldots, U_1,
 U_{n} U_{n}, U_{n} U_{n-1}, \ldots, U_1 U_1
 \right)
 .
 \label{input.poly}
\end{align}
We denote the generalized transition matrix as
\begin{align}
 {\bf \Xi}
  &=
\left(
 {\bf A} \;
 {\bf B}
\right).
\end{align}
The idea of DMD is to estimate the transition matrix from the
data.
This is done by taking pseudo inverse of the state variables as
\begin{align}
 \hat{ {\bf \Xi}}
 &=
 {\bf Y}_{k+1} \cdot
 \begin{pmatrix}
 {\bf Y}_{k} \\
 {\bf U}_{k}
\end{pmatrix}
 ^{\dagger}
 .
\end{align}
Here, $\mathbf{M} ^{\dagger}$ is the pseudoinverse of the matrix $\mathbf{M}$.
This is nothing but a least-square estimation for the cost function of l.h.s minus r.h.s of \eqref{narma10.exDMD}.
We may include the Tikhonov regularization term.

Note that for the extended DMD\cite{Li2017} and the higher-order DMD\cite{LeClainche:2017}, the transition matrix ${\bf \Xi}$ is further decomposed into characteristic modes associated with its eigenvalues.
The decomposition gives us a dimensional reduction of the system.
The estimation of the transition matrix is also called nonlinear system identification, particularly, nonlinear autoregression with exogenous inputs (NARX).
In this work, we focus on the estimation of the input-output relationship, and do not discuss the dimensional reduction.
For time-series prediction, we estimate the function $Y_{n+1} = f( Y_n,Y_{n-1}, \ldots,Y_1 )$, and we do not need the input $\mathbf{U}_n$ in \eqref{narma10.exDMD}.
Even in this case, we may consider a similar estimation of ${\bf \Xi}$ (in fact, $\mathbf{A}$).
This estimation is the method used in the next-generation RC\cite{Gauthier:2021}.

The second method is based on the Volterra series of the state variable
${\bf Y}_n$ by the input ${\bf U}_n$.
 In this method, we assume that the state variable is independent of
 its initial condition.
 Then, we may express the state variable as
\begin{align}
 {\bf Y}_n
 &=
 {\bf G} \cdot {\bf U}_n
 .
\end{align}
Note that $ {\bf U}_n$ includes the input and its polynomials with a time
delay as in \eqref{input.poly}.
Similar to the first method, we estimate ${\bf G}$ by
\begin{align}
 \hat{\bf G}
 &=
 {\bf Y}_t
 \cdot
 {\bf U}_t^{\dagger}
 .
\end{align}
The estimated $ \hat{\bf G}$ gives us information on which time delay
and nonlinearity dominate the state variable.

\begin{figure}[htbp]
\begin{center}
\includegraphics[width=0.80\textwidth]{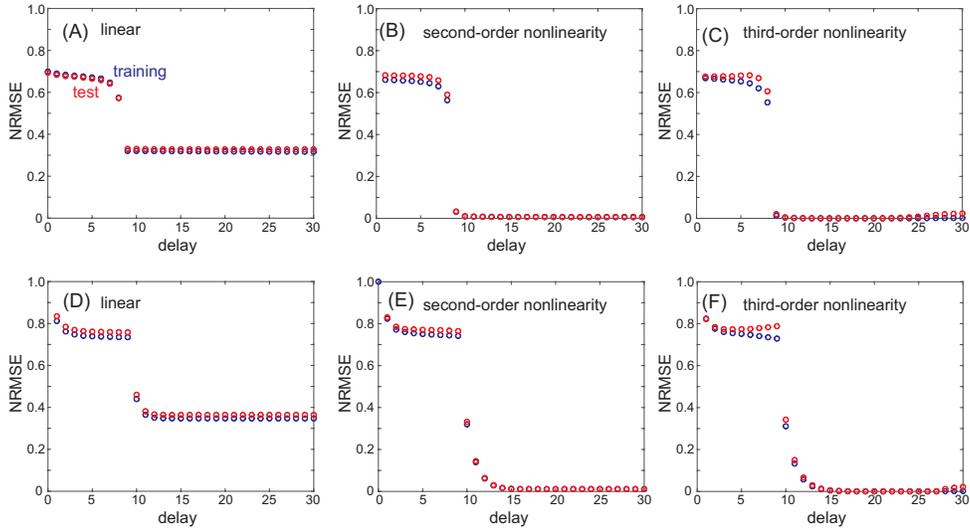}
 \caption{
 (A-C) the estimation based on the extended DMD, (D-F) the estimation
 based on the Volterra series.
 The dictionary of each case is (A,D) first-order (linear) delay terms, (B,E) up to second-order delay terms, and (C,F) up to third-order delay terms.
\label{fig.narma10.estimation}
}
\end{center}
\end{figure}

The results of the two estimation methods are shown in Fig.~\ref{fig.narma10.estimation}.
Both approaches suggest that memory of $\approx 10$ steps is enough to get
high performance, and further memory does not improve the error.
The second-order nonlinear term shows a reasonably small NRMSE of
$\approx 0.01$.
Including the third-order nonlinearity improves the error, but there is
a sign of overfitting at a longer delay because the number of the state
variables is too large.
It should also be noted that even with the linear terms, the NRMSE
becomes $\approx 0.35$.
This result implies that although $\mathrm{NRMSE} \approx 0.35$ is often considered good performance, nonlinearity of the data is not learned at the error of this order.

\subsection{The MC and IPC tasks as Volterra series for linear and nonlinear readout}\label{sec.mcipc.volterra}

In \eqref{m.GreenFunction} and \eqref{m2.GreenFunction} in the main text, we show that  the magnetization at the input region is expressed by the response
function.
The magnetization at the time $t_n$ corresponding to the input $U_n$ at the $n$ step is expressed as
\begin{align}
 m (t_n)
 &=
 a_n U_n +  a_{n-1} U_{n-1} + \cdots
 \label{m.volterra}
 ,
\end{align}
where the coefficients $a_n$ can be computed from the response function.
We first consider the linear case, but we will generalize the expression
for the nonlinear case.
 Because we use virtual nodes, the input $U_n$ at the step $n$ continues during the time period $t \in [t_n, t_{n+1})$ discretized by $N_v$ steps as $(t_{n,1},t_{n,2}, \ldots, t_{n,N_v})$, and is
 multiplied by the filter of the binary noise (see Fig.\ref{fig2} and Methods in the main text).
Therefore, the magnetization is expressed by the response functions $G (t-t')$
is formally expressed as
 \begin{align}
 m (t_n)
 =&
 \sum_i^{N_p}
 \left[
  \left(
G(0) 
+ G(\theta ) 
+ \cdots G(\theta (N_v-1))
  \right)
  \sigma_i (t_n)
\right.  
\nonumber \\
& \left.
  +  
  \left(
G(\theta N_v) 
+ G(\theta (N_v+1)) 
+ \cdots G(\theta (2N_v-1)) 
  \right) \sigma_i (t_{n-1})
 \right.
  \nonumber \\
&   \left.
 + \cdots
  \right]
  ,
  \label{m.Green.volterra}
\end{align}
where $\sigma_i (t_n) \propto U_n$ is the non-dimensionalized current injection at the time $t_n$ at the $i$th physical node, which is proportional to $U_n$.
Therefore, \eqref{m.Green.volterra} results in the expression of (\ref{m.volterra}).
Our input is taken from a uniform random distribution.
Therefore, the inner product of the reservoir state, which is nothing
but magnetization, and (delayed) input to learn MC is
\begin{align}
 \mean<m(t_n), U_n>
 &=
 \sum_{n=1}^T
 m (t_n) U_n 
 = a_n \mean<U_n^2>
 + \mathcal{O}(1/T)
 .
\end{align}
Similarly, the variance of the magnetization is equal to the variance of
the input with the coefficient associated with $m(t_n)$.

We may express the MC and IPC tasks in a matrix form as
\begin{align}
 \tilde{\bf S}
 &\approx
 {\bf W}
 \cdot {\bf G}
 \cdot \left(
 {\bf S}
 \circ {\bf W}_{\rm in}
 \right)
 .
 \label{volterra.SWAS}
\end{align}
Here, ${\bf S}$ is the matrix associated with the original input, and $ \tilde{\bf S}$ is the delayed one.
The output weight is denoted by ${\bf W}$, and ${\bf W}_{\rm in}$ is the matrix associated with the mask of binary noise.
The goal of MC and IPC tasks is to approximate the delayed input $\tilde{\bf S}$ by the reservoir states ${\bf G} \cdot {\bf S}$.
Here, the reservoir states are expressed by the response function ${\bf G}$ and input denoted by ${\bf S}$.
We define delayed input $\tilde{\bf S}  \in \mathbb{R}^{K \times T}$
\begin{align}
 \tilde{\bf S}
 &=
 \begin{pmatrix}
 U_{n} & U_{n+1} & U_{n+2} & \cdots \\
  U_{n-1} & U_{n}  & U_{n+1} & \cdots \\
 U_{n-2} & U_{n-1} & U_{n} & \cdots \\
   \vdots & \vdots  & \vdots  & \vdots  
 \end{pmatrix}
 .
\end{align}
Here, $T$ is the number of the time series, and $K$ is the total length
of the delay that we consider.
The $i$th row shows the $i-1$ delayed time series.
The input ${\bf S}  \in \mathbb{R}^{TN_v \times T}$ to compute the reservoir states are expressed as
\begin{align}
 \bf S
 &=
 \begin{pmatrix}
  \sigma (t_{n}) & \sigma (t_{n+1}) & \sigma (t_{n+2}) & \cdots \\
  \vdots & \vdots  & \vdots  & \vdots  \\
    \sigma (t_{n}) & \sigma (t_{n+1}) & \sigma (t_{n+2}) & \cdots \\
  \sigma (t_{n-1}) & \sigma (t_{n}) & \sigma (t_{n+1}) & \cdots \\
      \vdots & \vdots  & \vdots  & \vdots  \\
  \sigma (t_{n-2}) & \sigma (t_{n-1}) & \sigma (t_{n}) & \cdots \\
      \vdots & \vdots  & \vdots  & \vdots  
 \end{pmatrix}
 .
\end{align}
Note that $\sigma (t_n) \propto U_n$ upto constant.
Due to time multiplexing, each row is repeated $N_v$ times, and then the
time series is delayed in the next row.
After multiplying the input filter $\mathbf{W}_{\rm in}$, the input is fed into the response function.
The input filter $\mathbf{W}_{\rm in} \in \mathbb{R}^{T N_v \times T}$ is a stack of constant row vectors with the length $T$.
The $N_v$ different realizations of row vectors are taken from binary noise, and then the resulting $N_v \times T$ matrix is repeated $T$ times in the row direction.  
This input is multiplied by the coefficients of the Volterra series ${\bf G} \in \mathbb{R}^{N \times TN_v}$
\begin{align}
 {\bf G}
 &=
 \begin{pmatrix}
  G^{(1)} (0) &  \cdots & 
  G^{(1)} (\theta (N_v-1))
  & G^{(1)} (\theta N_v) & \cdots 
  & G^{(1)} (\theta (2N_v-1)) & \cdots \\
    G^{(2)} (0) &  \cdots & 
    G^{(2)} (\theta (N_v-1))
  & G^{(2)} (\theta N_v) & \cdots & 
  G^{(2)} (\theta (2N_v-1)) & \cdots \\
  \vdots &   \vdots &   \vdots &   \vdots & \vdots &   \vdots &   \vdots  \\
  G^{(N)} (0) &  \cdots & 
  G^{(N)} (\theta (N_v-1))
  & G^{(N)} (\theta N_v) & \cdots & 
  G^{(N)} (\theta (2N_v-1)) & \cdots 
 \end{pmatrix}
 \label{volterra.A}
\end{align}

(\ref{volterra.SWAS})  implies that by choosing the appropriate ${\bf W}$, we
can get a canonical form of ${\bf G}$.
If the canonical form has $N \times N$ identity matrix in the left part
of ${\bf W} \cdot {\bf G}$, then the reservoir reproduces the time
series up to $N-1$ delay.
This means that the rank of the matrix $G$, or the number of independent
rows, is the maximum number of steps of the delay.
This is consistent with the known fact that MC is bounded by the number
of independent components of reservoir variables\cite{Jaeger:2002}.

Next we extend the Volterra series of the magnetization, including
nonlinear terms.
The magnetization is expressed as
\begin{align}
 m (t_n)
 &=
 a_n \sigma(t_n) + 
 a_{n-1} \sigma (t_{n-1}) + \cdots
 +  a_{n,n} \sigma (t_n) \sigma (t_n) +
 a_{n,n-1} \sigma (t_n) \sigma (t_{n-1}) 
 + \cdots
 .
 \label{m.volterra.nonl}
\end{align}
The delayed input $\tilde{\bf S}$ is rewritten as
\begin{align}
 \tilde{\bf S}
 &=
 \begin{pmatrix}
  U_{n} & U_{n+1} & U_{n+2} & \cdots \\
  U_{n-1} & U_{n} & U_{n+1} & \cdots \\
  \vdots & \vdots  & \vdots  & \vdots\\
  U_{n} U_{n} & 
  U_{n+1} U_{n+1} &
  U_{n+2} U_{n+2} & \cdots \\
    U_{n} U_{n-1} & 
    U_{n+1} U_{n} & 
    U_{n+2} U_{n+1} & \cdots \\  
     \vdots & \vdots  & \vdots  & \vdots  
 \end{pmatrix}
 .
\end{align}
The matrix $\tilde{\bf S}$ contains all the nonlinear combinations of the input
series $(U_n, U_{n+1}, \cdots)$.
Accordingly, we should modify ${\bf S}$ and also ${\bf G}$ to include the nonlinear response
functions.
Note that to guarantee the orthogonality, Legendre polynomials (or other
orthogonal polynomials) should be used instead of polynomials in powers. 
Nevertheless, up to the second order of nonlinearity, which is relevant
to consider the performance of NARMA10 (see Sec.~\ref{sec.narma10}), the difference
is only in the constant terms ($P_2(x)=x^2 -\frac{1}{2}$).
Because we subtract the mean value of the time series of all the input,
output, and reservoir states, these constant terms do not change our
conclusion.
With nonlinear terms, (\ref{volterra.A}) is extended as ${\bf G}= ({\bf
G}_{\rm lin}, {\bf G}_{\rm nonl})$.
Still, the rank of the matrix remains $N$ at most.
This is the reason why the total sum of IPC, including all the linear and
nonlinear delays, is bounded by the number of independent reservoir
variables.
When ${\bf G}_{\rm nonl}={\bf 0}$, the reservoir can memorize only the
linear delay terms, but MC can be maximized to be $N$.
On the other hand, when ${\bf G}_{\rm nonl} \neq {\bf 0}$, it is
possible that MC is less than $N$, but the reservoir may have finite IPC.

When the readout is nonlinear, we use the reservoir state variable as
 \begin{align}
  {\bf X}
  &=
 \begin{pmatrix}
  {\bf M} \\
  {\bf M} \circ {\bf M}
 \end{pmatrix}
 ,
 \end{align}
 where $\circ$ is the Hadamard product.
 If ${\bf M}$ is linear in the input, ${\bf G}$ has a structure of
 \begin{align}
  {\bf G}
  &=
  \begin{pmatrix}
   {\bf G}_{\rm lin} & {\bf 0}\\
  {\bf 0} &     {\bf G}_{\rm nonlin}
  \end{pmatrix}
  .
 \end{align}
 In this case, $
  {\rm rank} ({\bf G})
  = {\rm rank} ({\bf G}_{\rm lin})
+ {\rm rank} ({\bf G}_{\rm nonlin})$.


\section{Learning with multiple variables}\label{sec.mumaxmxmz}

\begin{figure}[h]%
\centering
\includegraphics[width=0.75\textwidth]{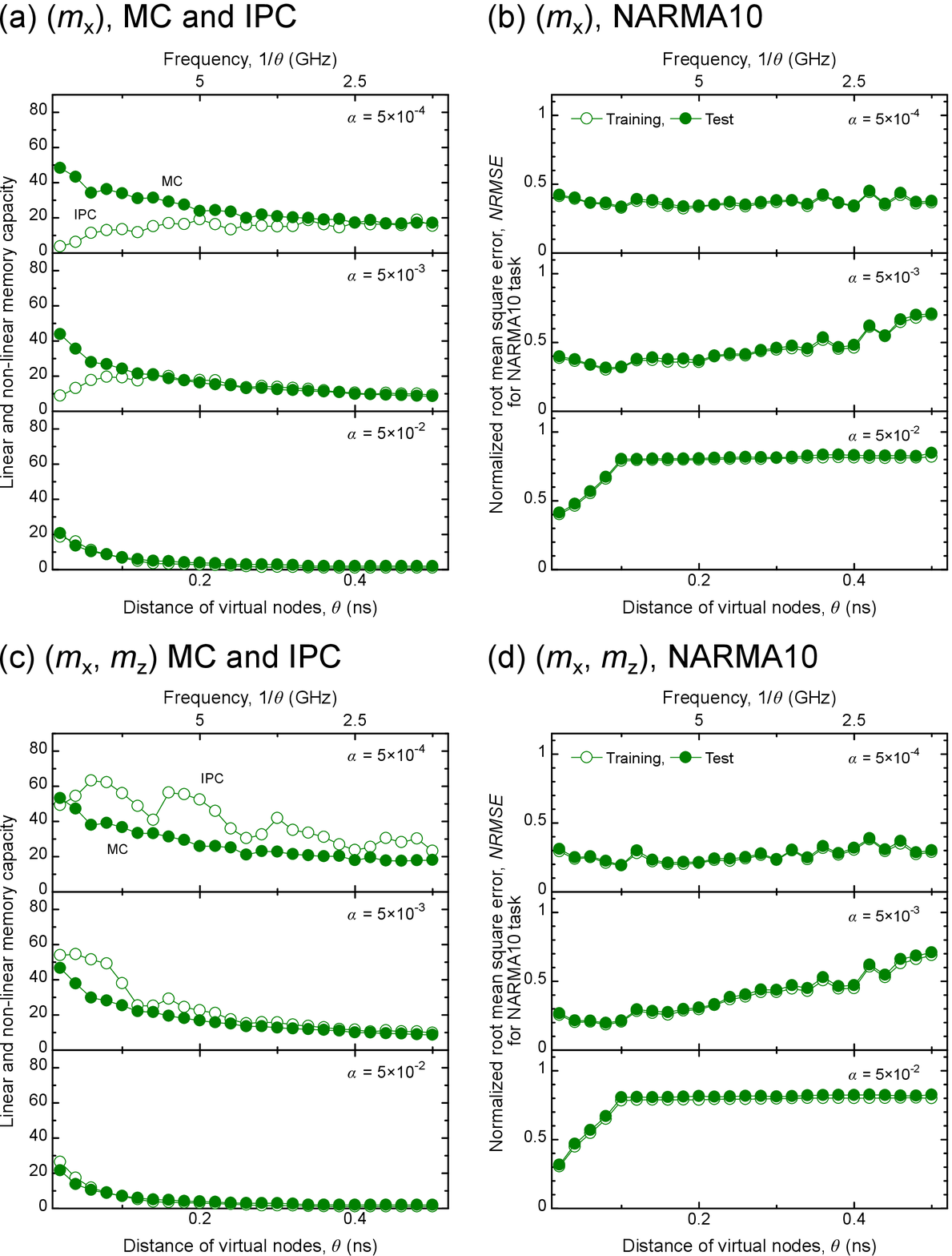}
\caption{ Reservoir computing with various parameter combinations obtained using micromagnetic Mumax3 simulation. Linear memory capacity, MC and nonlinear memory capacity, IPC plotted as a function of $\theta $ obtained using linear $m_x$ output only (a) and using {$m_x, m_z$} (c). Normalized root mean square error, NRMSE for NARMA10 task plotted as a function of $\theta $ obtained using linear $m_x$ output only (b) and using {$m_x, m_z$} (d).
}\label{figSmxmz}
\end{figure}

In the main text, we use only $m_x$ for the readout as in \eqref{readout.mx}-\eqref{nonl.readout}.
The readout is nonlinear and has both the information of $m_x$ and $m_x^2$.
In this section, we consider the linear readout, but use both $m_x$ and $m_z$ for the output in micromagnetic simulations.
We begin with the linear readout only with $m_x$.
The results of the MC and IPC tasks are shown in Fig.~\ref{figSmxmz}(a,b).
We obtain a similar performance for the MC task with the result in the main text (Fig.~\ref{fig3}).
On the other hand, the performance for the IPC task in Fig.~\ref{figSmxmz}(a) is significantly poorer than the result in Fig.~\ref{fig3}(a).
This result demonstrates that the linear readout only with $m_x$ does not learn the nonlinearity effectively.
Note that in the theoretical model with the response function, the IPC is exactly zero when we use the linear readout only with $m_x$.
The discrepancy arises from the expansion \eqref{m.expansion} around $\mathbf{m}^{0}=(0,0,1)$ in the main text.
Strictly speaking, the expansion should be made around $\mathbf{m}^{0}$ under the constant input $\mean<\sigma>$ averaged over time at the input nanocontact.
This reference state is inhomogeneous in space, and is hard to compute analytically.
Due to this effect, $m_x$ in the micromagnetic simulations contain small nonlinearity.

Next, we consider the linear readout with $m_x$ and $m_z$.
As seen in Fig.~\ref{figSmxmz}(c,d), $m_z$ carries nonlinear information, and enhances the IPC and learning performance of NARMA10 compared with linear readout only with $m_x$ (Fig.~\ref{figSmxmz} (a,b)).
The performance is $\mathrm{IPC} \approx 60$ under $\alpha = 5\times 10^{-4}$, which is comparable value with the results in the main text (Fig.~\ref{fig3}(a,c)) where the readout is $(m_x, m_x^2)$.
Also, high performance for NARMA10 task, $\mathrm{NRMSE} \approx 0.2$, can be obtained using variables $( m_x, m_z )$.
These results show that adding $m_z$ into the readout has a similar effect to adding $m_x^2$.

Similarity between $m_x^2$ and $m_z$ can be understood by using the theoretical formula with the response function in the main text.
We continue the expansion \eqref{m.expansion} at the second order, and obtain
\begin{align}
    \partial_t \mathbf{m}^{(2)} (\mathbf{x},t)
    =&
    - \mathbf{m}^{(1)} \times \Delta \mathbf{m}^{(1)}
    - \alpha \mathbf{m}^{(1)}  \times
    \left(
    \left(
    \tilde{h}\mathbf{m}^{(1)} - \Delta \mathbf{m}^{(1)}
    \right)
    \times \mathbf{e}_z
    \right)
    \nonumber \\
    &+
    \sigma (\mathbf{x},t) \mathbf{m}^{(1)} \times \mathbf{e}_y
    .
\end{align}
This result suggests that $\mathbf{m}^{(2)}$ contains only the $z$ component, and is slaved by $\mathbf{m}^{(1)}$, which does not have $z$ component.
Therefore, $m_z^{(2)}$ can be computed as
\begin{align}
    m_z^{(2)} (\mathbf{x},t) 
    &= 
    - \frac{1}{2} \left( 
    (m_x^{(1)})^2 + (m_y^{(1)})^2
    \right)
    .
\end{align}
Because $m_x$ and $m_y$ carry similar information, $m_z$ in the readout has a similar effect with $m_x^2$ in the readout.


\section{Speed of propagating spin wave using dipole interaction}
\label{sec.speed}

Propagating spin wave when magnetization is pointing along film normal is called magneto-static forward volume mode, and its dispersion relation can be described by the following equation\cite{Hillebrands2007}.
\begin{align}
\omega (k) = \gamma \mu _0 \sqrt{\left(H_0-M_s\right) \left( H_0-M_s \frac{1-e^{-kd}}{kd}\right)}.    
\end{align}
Then, one can obtain the group velocity at $k$ $\sim $ 0 as,
\begin{align}
v_g=\frac{d\omega }{dk} (k=0) = \frac{\gamma \mu_0 M_s d}{4}. \label{eq:vg}
\end{align}
In the magneto-static spin wave driven by dipole interaction, group velocity is proportional to both $M_s$ and $d$.
$v_g$ $\sim $ 200 m/s is obtained when the following parameters are used: $\mu _0 H$ = 1.5 T, $M_s$ = 1.0 $\times $ 10$^6$ A/m, $d$ = 4 nm.
The same estimation is used for calculating the speed of information propagation for spin reservoirs in Refs. \cite{Nakane:2021} and \cite{Dale:2021}, which are used to plot Fig.~\ref{fig6} in the main text. 


\section{Details of reservoir computing scaling compared with literature}
\label{sec.literature}

In this section, details of Fig.~\ref{fig6} shown in the main text are described.
MC and NRMSE for NARMA10 tasks using photonic and spintronic RC are
reported in Refs. \cite{Paquot2012, Duport2012, Dejonckheere2014,
Vinckier2015, Duport2016, Takano:2018, Sugano2020} for photonic RC and
\cite{Furuta:2018, Tsunegi:2019, Kanao2019, Akashi2020, Nakane:2021,
Watt2021, Dale:2021, Lee2022} for spintronic RC.
Table \ref{tableS1} and \ref{tableS2} shows reports of MC for photonic
and spintronic RC with different length scales, which are plotted in Fig.~\ref{fig6} in the main text.

\begin{table}[h]
\begin{center}
\begin{minipage}{\textwidth}
\caption{Report of photonic RC with different length scales used in Fig.~\ref{fig6} in the main text}\label{tableS1}
\begin{tabular*}{\textwidth}{@{\extracolsep{\fill}}lccccc@{\extracolsep{\fill}}}
\toprule%
Reports & Length, $L$ & Time interval, $\tau _0$ & $v\tau _0$ & $N$ & $MC$  \\
\midrule
Duport {\it et al.}\cite{Duport2012}  & 1.6 km & 8 $\mu $s & 2.4 km & 50 & 21 \\
Dejonckheere {\it et al.}\cite{Dejonckheere2014}  & 1.6 km & 8 $\mu $s  & 2.4 km  & 50 & 37  \\
Vincker {\it et al.}\cite{Vinckier2015}  & 230 m & 1.1 $\mu $s  & 340 m  & 50 & 21 \\
Takano {\it et al.}\cite{Takano:2018}  & 11 mm & 200 ps  & 60 mm  & 31 & 1.5  \\
Sugano {\it et al.}\cite{Sugano2020}  & 10 mm & 240 ps  & 72 mm & 240 & 10 \\
\bottomrule 
\end{tabular*}
\footnotetext{Note: speed of light, $v$ = 3 $\times $ 10$^8$ m/s is used.}
\end{minipage}
\end{center}
\end{table}

\begin{table}[h]
\begin{center}
\begin{minipage}{\textwidth}
\caption{Report of spin reservoirs with different length scales used in Fig.~\ref{fig6} in the main text}\label{tableS2}
\begin{tabular*}{\textwidth}{@{\extracolsep{\fill}}lcccccc@{\extracolsep{\fill}}}
\toprule%
Reports & $L$ &$\tau _0$ & $v$ & $v\tau _0$ & $N$ & $MC$  \\
\midrule
Nakane {\it et al.}\cite{Nakane:2021}  & 5 $\mu $m & 2 ns & 2.4 km/s & 4.8 $\mu $m & 72 & 21 \\
Dale {\it et al.}\cite{Dale:2021} & 50 nm & 10 ps & 200 m/s & 2 nm  & 100 & 35  \\
This work  & 500 nm & 1.6 ns & 200 m/s & 320 nm & 64 & 26 \\
\bottomrule 
\end{tabular*}
\footnotetext{Note: $v$ is calculated based on magneto-static spin wave using Eq. \ref{eq:vg}.}
\end{minipage}
\end{center}
\end{table}


\section{Other data}
\label{sec.other}

\subsection{$N_v$ and $N_p$ dependence of performance}

\begin{figure}[h]%
\centering
\includegraphics[width=0.95\textwidth]{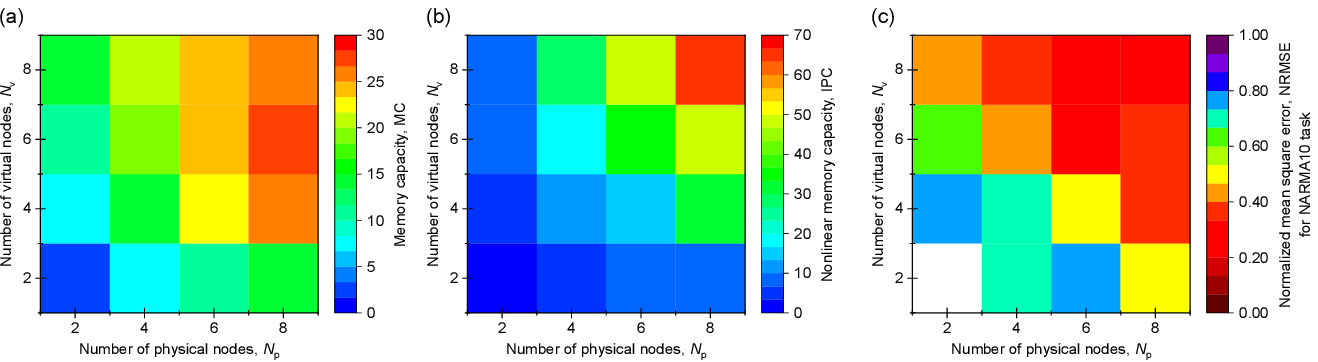}
\caption{
(a) Memory capacity, MC (b) Nonlinear memory capacity, IPC and (c) Normalized root mean square error, NRMSE for NARMA10 task plotted as a function of the number of virtual and physical nodes. The parameters used in the simulation are $\alpha = 5\times 10^{-4}, \theta = 0.2 $ ns.
}\label{figS3}
\end{figure}

Fig.~\ref{figS3} shows $N_v$ and $N_p$ dependencies of MC, IPC and NRMSE for NARMA10 task.
As $N_v$ and $N_p$ are increased, MC and IPC increase.
Then, NARMA10 prediction task becomes better with increasing $N_v$ and $N_p$.
MC and NRMSE for NARMA10 with different $N_p$ with fixed $N_v$ = 8 are compared with other reservoirs shown in Fig.~\ref{fig7} in the main text. 

\subsection{exchange interaction}

\begin{figure}[h]%
\centering
\includegraphics[width=0.98\textwidth]{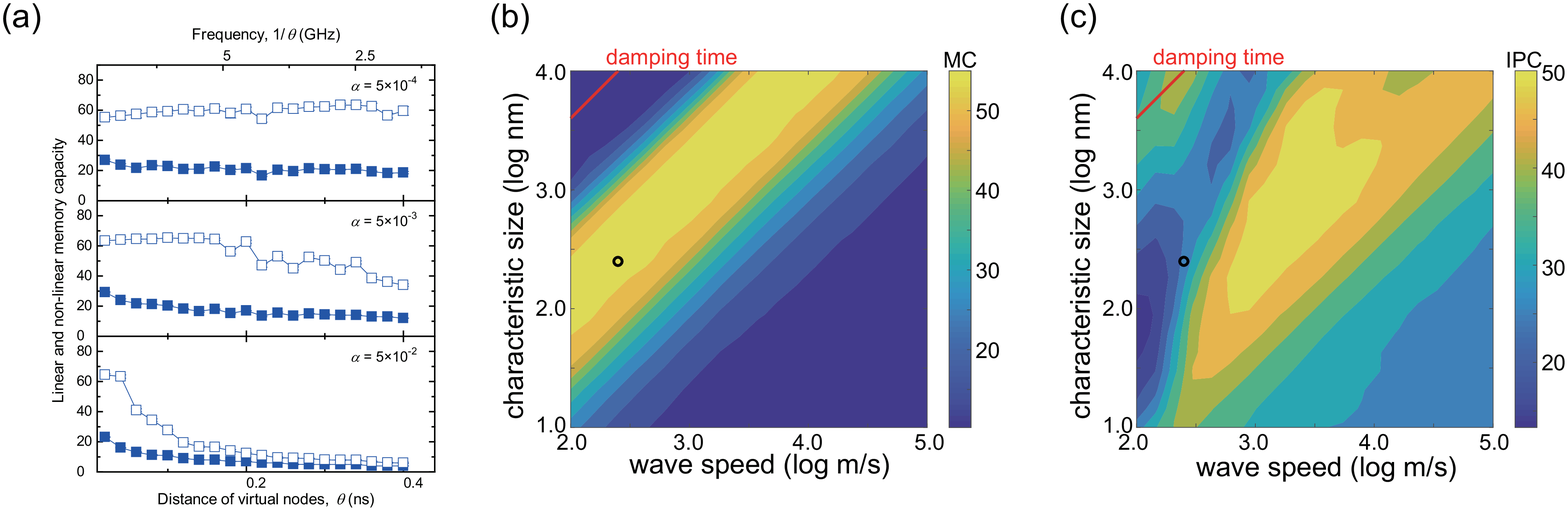}
\caption{(a) Memory capacity, MC (solid symbols) and nonlinear memory capacity, IPC (open symbols) obtained using the response function method for exchange interaction plotted as a function of $\theta $ with different damping parameters $\alpha $. (b) MC and (c) IPC plotted as a function of characteristic size and wave speed.
}\label{figSexchange}
\end{figure}

In the main text, we use the dipole interaction to compute the response function as \eqref{Giidipole} and \eqref{Gijdipole}.
In this section, we show the result using the exchange interaction shown in \eqref{Giiexchange} and \eqref{Gijexchange}.
Figure~\ref{figSexchange} shows the results.

\nocite{*}
\bibliography{spinwaveRC}
\bibliographystyle{ScienceAdvances}

\end{document}